\documentstyle[localsprocl]{article}

\newcommand{\myftnote}[1]{\footnote{#1}}
\newcounter{exer}[section]

\newif\iffn\fnfalse

\input{epsf}
\newcommand {\realR} {{\Bbb R}}
\newcommand {\intZ} {{\Bbb Z}}

\def\noj#1,#2,{{\bf #1} (19#2)\ }
\def\jou#1,#2,#3,{{\sl #1\/ }{\bf #2} (19#3)\ }
\def\ann#1,#2,{{\sl Ann.\ Physics\/ }{\bf #1} (19#2)\ }
\def\cmp#1,#2,{{\sl Comm.\ Math.\ Phys.\/ }{\bf #1} (19#2)\ }
\def\invm#1,#2,{{\sl Invent.\ Math.\/ }{\bf #1} (19#2)\ }
\def\cq#1,#2,{{\sl Class.\ Quantum Grav.\/ }{\bf #1} (19#2)\ }
\def\cqg#1,#2,{{\sl Class.\ Quantum Grav.\/ }{\bf #1} (19#2)\ }
\def\ijmp#1,#2,{{\sl Int.\ J.\ Mod.\ Phys.\/ }{\bf A#1} (19#2)\ }
\def\jmphy#1,#2,{{\sl J.\ Geom.\ Phys.\/ }{\bf #1} (19#2)\ }
\def\jams#1,#2,{{\sl J.\ Amer.\ Math.\ Soc.\/ }{\bf #1} (19#2)\ }
\def\grg#1,#2,{{\sl Gen.\ Rel.\ Grav.\/ }{\bf #1} (19#2)\ }
\def\mpl#1,#2,{{\sl Mod.\ Phys.\ Lett.\/ }{\bf A#1} (19#2)\ }
\def\nc#1,#2,{{\sl Nuovo Cim.\/ }{\bf #1} (19#2)\ }
\def\np#1,#2,{{\sl Nucl.\ Phys.\/ }{\bf B#1} (19#2)\ }
\def\pl#1,#2,{{\sl Phys.\ Lett.\/ }{\bf #1B} (19#2)\ }
\def\pla#1,#2,{{\sl Phys.\ Lett.\/ }{\bf #1A} (19#2)\ }
\def\pr#1,#2,{{\sl Phys.\ Rev.\/ }{\bf #1} (19#2)\ }
\def\prd#1,#2,{{\sl Phys.\ Rev.\/ }{\bf D#1} (19#2)\ }
\def\prl#1,#2,{{\sl Phys.\ Rev.\ Lett.\/ }{\bf #1} (19#2)\ }
\def\prp#1,#2,{{\sl Phys.\ Rept.\/ }{\bf #1C} (19#2)\ }
\def\ptp#1,#2,{{\sl Prog.\ Theor.\ Phys.\/ }{\bf #1} (19#2)\ }
\def\ptpsup#1,#2,{{\sl Prog.\ Theor.\ Phys.\/ Suppl.\/ }{\bf #1} (19#2)\ }
\def\rmp#1,#2,{{\sl Rev.\ Mod.\ Phys.\/ }{\bf #1} (19#2)\ }
\def\yadfiz#1,#2,#3[#4,#5]{{\sl Yad.\ Fiz.\/ }{\bf #1} (19#2) #3%
\ [{\sl Sov.\ J.\ Nucl.\ Phys.\/ }{\bf #4} (19#2) #5]}
\def\zh#1,#2,#3[#4,#5]{{\sl Zh.\ Exp.\ Theor.\ Fiz.\/ }{\bf #1} (19#2) #3%
\ [{\sl Sov.\ Phys.\ JETP\/ }{\bf #4} (19#2) #5]}
\def\beq{\begin{equation}}
\def\eeq{\end{equation}}
\def\beqar{\begin{eqnarray}}
\def\eeqar{\end{eqnarray}}

\def\nfrac#1#2{{\displaystyle{\vphantom1\smash{\lower.5ex\hbox{\small$#1$}}%
	\over\vphantom1\smash{\raise.25ex\hbox{\small$#2$}}}}}

\def\to{\rightarrow}
\def\implies{\Rightarrow}

\def\lae{\mathrel{\mathop{\smash{\lower .5 ex \hbox{$\stackrel<\sim$}}}}}
\def\lae{\mathrel{\mathop{\smash{\lower .5 ex \hbox{$\stackrel>\sim$}}}}}

\newfont{\bbbfont}{msbm10}
\newfont{\smallbbbfont}{msbm7}
\newfont{\tinybbbfont}{msbm5}
\newcommand{\Bbb}[1]{
      \mathchoice{\mbox{\bbbfont #1}}{\mbox{\bbbfont #1}}
      {\mbox{\smallbbbfont #1}}{\mbox{\tinybbbfont #1}}}
\def\ket#1{\left| #1 \right\rangle}

\def\pa{\partial}
\def\pb{\bar\pa}
\def\na{\nabla}
\def\Tr{{\rm Tr \,}}
\def\l:{\mathopen{:}\,}
\def\r:{\,\mathclose{:}}

\def\half{\nfrac12}

\newcommand{\beqn}{\begin{equation}}
\newcommand{\eeqn}{\end{equation}}
\newcommand{\beqnarray}{\begin{eqnarray}}
\newcommand{\eeqnarray}{\end{eqnarray}}

\newcommand{\Psibar}{\bar{\Psi}}

\newcommand \ncom {\newcommand}

\newcommand {\nono} {\nonumber \\} 

\newcommand {\bear} [1] {\begin {array} {#1}}
\newcommand {\ear} {\end {array}}

\newcommand {\myI} [1] {\int \! #1 \,}

\newcommand {\beqarn} {\begin{eqnarray*}}
\newcommand {\eeqarn} {\end{eqnarray*}}

\newcommand {\sepe} {\;\;\;\;\;\;\;\;}
\newcommand	{\come} {\;\;\;\;}

\newcommand {\psibar} {\bar {\psi}}
\newcommand {\zbar} {{\bar z}}
\newcommand {\zb} {\zbar}

\newcommand {\Cg} {{ g}}
\newcommand {\CG} {{\cal G}}

\newcommand {\quat} {\frac 1 4}

\newcommand {\qbar} {{\bar q}}
\newcommand {\slZ} {{\mbox {$\mrm {SL} (2, \intZ)$}}}
\newcommand	{\abs}	[1] {{\left| #1 \right|}}
\newcommand {\brac} [1]	{{\left\{	#1 \right\}}}
\newcommand	{\paren} [1] {{\left( #1 \right)}}
\newcommand	{\brak} [1] {{\left[ #1 \right]}}

\newcommand {\bs} {\backslash}

\newcommand {\mrm} [1] {\mathrm {#1}}

\newcommand {\im} {\mathrm {Im \,}}
\newcommand {\re} {\mathrm {Re \,}}

\newcommand {\imply} {\Rightarrow}
\newcommand {\bij} {\leftrightarrow}

\newcommand {\nod} [1] {\mbox {$:#1\!:$}}

\newcommand {\comm} [2] {\mbox {$\left[ #1, #2 \right]$}}
\newcommand {\acomm} [2] {\mbox {$\left\{ #1, #2 \right\}$}}

\newcommand {\column} [1] {{\paren {\bear {c} #1 \ear}}}

\newcommand {\zNS}	{Z_{\mrm{NS}}}
\newcommand {\zR}	{Z_{\mrm{R}}}

\newcommand {\Fas} [1] {{F^{\brac {#1}}}}
\newcommand {\hFas} [1] {{{\hat F}^{\brac {#1}}}}
\newcommand {\Aas} [1] {{A^{\brac {#1}}}}
\newcommand {\hAas} [1] {{{\hat A}^{\brac {#1}}}}

\newcommand {\vol} {{\mbox {$\mrm{vol}$}}}

\newcommand {\chap} {\S}

\newcommand {\tL} {{\tilde L}}

\newcommand {\dirac} {\not \!}

\newcommand {\myepsf} [1] 
	{\begin{figure} [htbp]
		\begin{center} 
			$\mbox{\epsfbox{#1}}$
		\end{center}
	\end{figure}}

\ncom \poincare {Poincare\'{e}}

\def\emph#1{{\em #1}}
\def\mathbf#1{{\bf #1}}
\def\mathrm#1{{\rm #1}}
\def\mathit#1{{\it #1}}

\begin{document}

\title{LECTURES ON PERTURBATIVE STRING THEORIES\footnote{Lectures
      given by H. Ooguri.}}

\author{HIROSI OOGURI and ZHENG YIN}

\address{Department of Physics,
University of California at Berkeley\\
366 Le\thinspace Conte Hall, Berkeley, CA 94720-7300, U.S.A.\\
and\\
Theoretical Physics Group, Mail Stop 50A--5101\\
Ernest Orlando Lawrence Berkeley National Laboratory, 
Berkeley, CA 94720, U.S.A.\\}

\maketitle\abstracts{
These lecture notes on String Theory 
constitute  an introductory course designed to acquaint 
the students with some basic facts of perturbative 
string theories.  They are intended as preparation for
the more advanced courses on non-perturbative aspects
of string theories in the school. 
The course consists 
of five lectures: 1. Bosonic String, 2. Toroidal Compactifications, 
3. Superstrings, 4. Heterotic Strings, and 5. Orbifold 
Compactifications. 
}

\tableofcontents

\newpage

\newpage

\section{Lecture One: Bosonic String}

It had been said that there are five different string
theories -- (1) open and closed superstrings (Type I), 
(2) non-chiral closed superstring (Type IIA),
(3) chiral closed superstring (Type IIB), 
(4) heterotic string with $E_8 \times E_8$ gauge symmetry,
and (5) heterotic string with {\it Spin}$(32)/\intZ_2$ gauge symmetry.
They are all formulated perturbatively as sums over
two-dimensional surfaces. It had been known for a long time 
(and as we will learn in this course) that (2) can be related to (3), and
(4) to (5), if we compacitify part 
of the target spacetime on $S^1$. These relations were discovered earlier 
since they hold in each order in the perturbative expansion of the
theories. During the last two years, 
it has become increasingly clear that in fact 
all these five string theories are related to each
other under various duality transformations. It seems
likely that there is something more fundamental, which
we may call {\it the theory}, and the five string theories describe various
asymptotic regions of it. 

One of the purposes of this year's TASI summer school is to 
guide students through this recent exciting development. 
It is hoped that students attending the school will someday
reveal what {\it the theory} is about. 
First, however, the students have to understand its 
five known asymptotic regions.
This is the purpose of this course.  
We will construct
and analyze four perturbative string theories, (2), (3), (4) and (5).
The type I  theory containing open superstring will not
be discussed here since it
will be covered in Polchinski's lecture in this
school. Due to the limited time, we cannot discuss computations
of higher-loop string amplitudes at all. This important topic
has been covered in previous TASI lectures, by Vafa \cite{vafa} 
and by D'Hoker \cite{dhoker}.  Due to the long history of the
string theory, we were unable to make a complete bibliography of
original papers. We apologize to numerous contributors to
the subject for the omission. For works before 1987, we refer 
to the bibliography of \cite{GSW}.

\subsection {Point Particle}

	Let us begin this lecture on string theory by recalling 
the relativistic action for a point particle moving in 
$D$-dimensional spacetime.
\beq	\label {ptact}
	S = - m \myI {d\sigma} \sqrt { - \dot {X}^2},
\eeq
where
\[ 
~~~~		\dot X ^\mu= \frac  {\pa X^\mu} {\pa \sigma},~~~
 \mu = 0,..., D-1.
\]
We are using the Minkowski metric $\eta_{\mu\nu}$ with signature 
$(-1,+1,\ldots,+1)$. 
It is an action defined over the worldline the particle traverses.
Its canonical momenta are given by 
\beq
	P_\mu = \frac{\pa L} {\pa \dot X^\mu} = m \frac  {\dot X_\mu} 
		{\sqrt{-\dot X ^2}}.
\eeq
However, they are not all independent but satisfy a constraint 
equation which is simply Einstein's relation between energy, 
momentum and mass.
\beq	\label {ptconstraint}
	P^2 = -m^2
\eeq
The constraint arises since
 (\ref {ptact}) is invariant under worldline 
reparametrization: 
$\sigma \rightarrow \sigma^\prime = f(\sigma)$.
This is a gauge symmetry that we naturally expect since changing 
the parametrization scheme of the worldline should have {\em no} 
physical effect at all.  It indicates the $X$'s and 
$P$'s are redundant as coordinates of the physical phase 
space.  We can eliminate one of the $X$'s by a choice 
of gauge.  For example, we can set $X^0 = \sigma$ so the 
worldline time $\sigma$
coincides with the physical time $X^0$.  Constraint (\ref {ptconstraint}) then 
tells us how to eliminate one of the $P$'s.    Upon quantization 
the constraint (\ref {ptconstraint}) then becomes the requirement 
that physical states and observables should be gauge 
invariant.

\subsection {Nambu-Goto Action}

We now formulate string theory as an analogue of (\ref {ptact}) 
on a two-dimensional {\em worldsheet}.  
The fields $X^\mu$ on the worldsheet $\Sigma$ define an embedding of $\Sigma$ 
in the $D$-dimensional spacetime.   The pull-back of the Minkowski metric
$\eta$ to the worldsheet is called an induced metric:
\beq	\label {indmet}
	\hat g_{a b} = \eta_{\mu \nu} \frac  {\pa X^\mu} {\pa \sigma^a} 
				\frac  {\pa X^\nu} {\pa \sigma^b}; \come
				 a, b = 0, 1; \come 
				 \hat g \equiv \det \hat
          g_{a b}.
\eeq
We then define the \emph {Nambu-Goto} string action as proportional to the area
of the worldsheet measured by the induced metric:
\beq		\label {nambugoto}
	S = \frac {- 1} {2 \pi \alpha^\prime} \int \!d^2\sigma \, 
		\sqrt{- \hat g}.
\eeq
Because there are only bosonic degrees of freedom involved, 
we call it bosonic string.  It can be shown 
that the dimensionful constant 
$T \equiv  1 / (2 \pi \alpha')$ 
gives the  tension of the string.

The coordinate $\sigma^0$ is the ``time'' on the string worldsheet, $\sigma^1$ 
is the ``space'' coordinate along the closed string.  In these lectures we 
only consider {\em orientable closed} strings, which means that
the worldsheet can be assigned a definite orientation.  
By a reparametrization, we can let $\sigma^1$ range from $0$ to $2\pi$.  
Thus the Nambu-Goto 
action describes the motion of a string in spacetime, and the worldsheet 
is the trajectory it sweeps out (fig. 1).
Unlike a particle, a string can have internal oscillations in 
addition to its center of mass motion.  They include  
oscillation both transverse and longitudinal to the string worldsheet.  
As we will now demonstrate, however, only the transverse oscillations 
are physical.

\myepsf{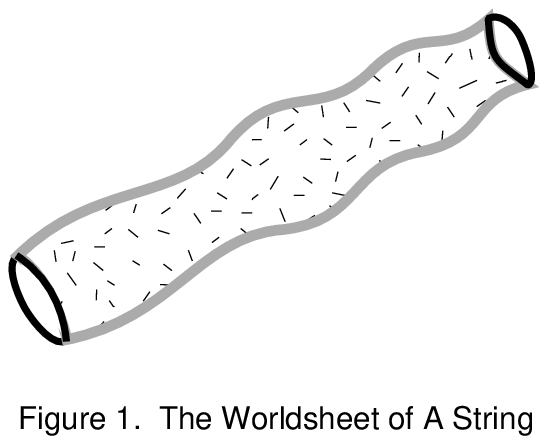}
	
	The canonical momentum densities of the Nambu-Goto action 
(\ref {nambugoto}) are given by 
\beq	\label {momentum}
	P_\mu (\sigma) = \frac {\delta L} {\delta\dot X^\mu(\sigma)} 
			= \frac 1 {2 \pi \alpha^\prime} \sqrt{- \hat g} 
				\; \hat g^{0 \alpha} \: \eta_{\mu \nu}
\pa_\alpha X^\nu.
\eeq
Since (\ref {nambugoto})  has reparametrization 
invariance on the worldsheet: 
$(\sigma^0, \sigma^1)$ $\rightarrow$ $(\sigma^{0 \prime}, \sigma^{1 \prime})$,
one naturally expects an analog of the constraint (\ref {ptconstraint}).  
Indeed one finds
\beqarn 
	P \cdot \pa_1 X &=& \frac 1 {2 \pi \alpha^\prime}  
							\sqrt{- \hat g} 
							\hat g^{0 \alpha}  \hat g_{\alpha 1} = 
						 \frac 1 {2 \pi \alpha^\prime}  
							\sqrt{- \hat g} \delta^0_1 = 0, \\
	P^2 &=& - (\frac 1 {2 \pi \alpha^\prime}) ^2 
				\hat g  \hat g^{0 0} 
			= - (\frac 1 {2 \pi \alpha^\prime}) ^2  g_{1 1}
			= - (\frac 1 {2 \pi \alpha^\prime}) ^2  (\pa_1 X)^2, 
\eeqarn
or
\beq	\label {constraint}
	\left[ P \pm \frac 1 {2 \pi \alpha^\prime}   (\pa_1 X)
\right]^2 = 0,
\eeq
known as the \emph {Virasoro constraints}.

	By using the reparametrization invariance, we can bring  
the induced metric on any {\em coordinate patch} of the worldsheet to be a 
multiple of the standard Lorentzian metric:
\beq	\label {cfgauge}
	\hat g_{ab} = \lambda \gamma_{ab}, \
	(\gamma_{ab}) = \left( {\bear {r r} -1 & 0 \\ 0 & 1 \ear} \right).
\eeq
Here $\lambda$ may be a function of $\sigma^a$. 
This is called the \emph {conformal gauge}.  Note that this choice 
of gauge does not break spacetime Lorentz invariance.  
In this gauge, the momentum densities $P_\mu$ are given by  
\beq		\label {cfmom}
	P_\mu = \frac 1 {2 \pi \alpha^\prime} \dot X_\mu, \come \dot X =
\pa_0 X ,
\eeq 
and the Virasoro constraints are
\beqarn
	\pa_0 X \cdot \pa_1 X &=& \hat g_{0 1} = 0 , \\
	(\pa_0 X)^2 + (\pa_1 X)^2 &=& \hat g_{0 0} + \hat g_{1 1} = 0 .
\eeqarn

In conformal gauge, there is still a residual gauge 
symmetry.   It is called \emph {conformal symmetry} because 
it only rescales the induced metric.  To exhibit it, 
define the light-cone coordinates 
$\sigma^\pm \equiv \sigma^0 \pm  \sigma^1$.  It is  not
difficult to show that a coordinate transformation preserving
the conformal gauge condition (\ref{cfgauge}) must be of
the form
\beq	\label {cftran}
	\sigma^+ \rightarrow \sigma^{+\prime} = f(\sigma^+), \sepe
	\sigma^- \rightarrow {\sigma^{-}}' = h(\sigma^-).
\eeq
In the light-cone coordinates,
\[  -(d\sigma^0)^2 + (d\sigma^1)^2 = - d\sigma^+ d \sigma^- . \]
Since
\[	d{\sigma'}^{+} = f' d\sigma^+, \sepe
    d{\sigma'}^{-} = h' d\sigma^-,	\]
(\ref{cfgauge}) is indeed preserved as
\[	 d{\sigma'}^+ d{\sigma'}^-	
	=  f' h' d\sigma^{+} d\sigma^{-}.	 
\]

The 
worldsheet of a freely propagating string 
clearly looks like a tube.  Choosing 
$L_n$ and $\tilde L_n$, the Fourier components of $f$ and $h$ 
respectively, as the generators of conformal transformation on 
a cylinder, it is not difficult to find their commutators:
\beqar	\label {cfalgebra}
	\comm {L_n} {L_m}\, &=& (n-m) L_{n+m}, \nono
	\comm {\tL_n} {\tL_m} &=& (n-m) \tL_{n+m}, \nono
	\comm {L_n} {\tL_m} &=& 0.	
\eeqar

We can completely fix this residual gauge symmetry, at the 
expense of spacetime Lorentz invariance.  
In the conformal gauge, the equation of motion for $X^\mu(\sigma)$ is 
\beq	\label {cfeom}
	(\pa_0^2 - \pa_1^2) X^\mu = 0,	
\eeq
and its general solution is
\[ 	X^\mu  = X_L^\mu (\sigma^+) + X_R^\mu (\sigma^-) .	\]
Define $X^+ \equiv X^0 + X^1$, $X^- \equiv X^0 - X^1$.  
Taking advantage of this residual gauge symmetry, we can always choose 
local coordinates so that:
\beq
	X^+ \equiv X^0 + X^1 = \alpha^\prime p^+ \sigma^0,
\eeq
where $p^+$ is a constant\myftnote
	{We keep $p^+$ rather than absorbing it into 
	$\sigma^0$ so as to preserve the canonical Poisson bracket.  
	In fact, since $p^+$ is a physical observable, we should not 
	be able to gauge it away.} which, by (\ref{momentum}), is
related to the  
momentum density as \(P^+ =  {p^+} / 2 \pi\). 
Substituting this into (\ref{constraint}),  we obtain 
\beq	\label {lcsol}	
	\pa_\pm X^- = \frac 1 {\alpha' p^+}
  \sum_{i = 2}^D \pa_\pm X^i \pa_\pm X^i.
\eeq
This determines $X^- = X^0 - X^1$ up to a constant of integration $x^-$.
Thus the gauge invariant information of a propagating string 
is given by $x^-$, 
$p^+$, and $X^i(\sigma^0, \sigma^1)$, $(i = 2, \ldots, D-1)$.
As in the case of point particles, the worldsheet 
reparametrization invariance removes all degrees of freedom along the 
light-cone directions except for their zero modes.  
Intuitively, one can understand 
this as saying that oscillations tangential to the string 
can be absorbed 
by a worldsheet reparametrization 
(fig. 2).

\myepsf {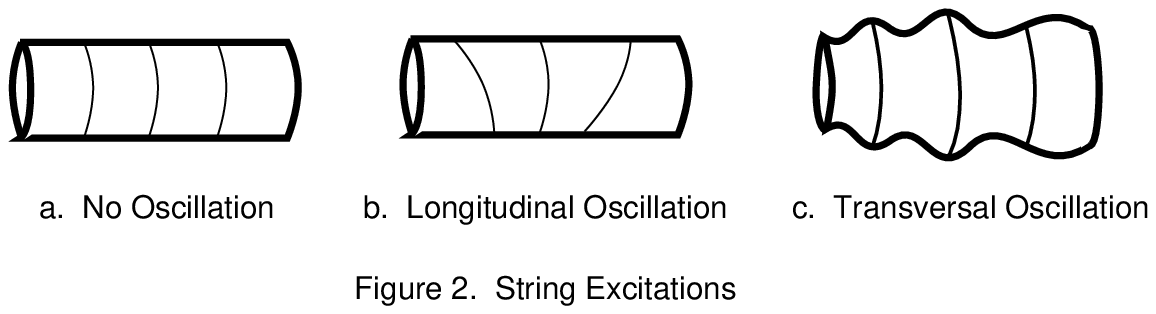}

On length scales much larger than the string scale $\sqrt {\alpha'}$, 
which is the typical size of string, 
low lying excitations of 
string look like point particles and should form unitary representations 
of the Poincar\'e group.  They are classified by the little group of their 
momenta in a certain Lorentz frame.
As we will see, the first excitation level 
of the string makes a rank 2 tensor 
representation of $SO(D-2)$, including the trace, traceless symmetric
and antisymmetric parts. The little groups 
for massless and massive particles in $(1,D-1)$ spacetime are 
$SO(D-2)$ and $SO(D-1)$ respectively.
Since the rank 2 tensor of $SO(D-2)$ alone cannot be made into
a representation of $SO(D-1)$, these excitations must correspond to 
massless particles.  According to Weinberg's theorem \cite {weinberg}, 
a massless rank 2 symmetric traceless tensor that is observed at low 
energy must describe graviton and implies general covariance. 
Hence a theory of closed strings must be, among other things, 
a theory of gravity.

\subsection {First Quantization of String}

	For point particles, there are two roads from classical physics 
to quantum physics.  The first quantization quantizes 
the worldline action and yields quantum mechanics 
(i.e. one-dimensional QFT) of the particles.  
The second quantization  quantizes their spacetime action and 
yields a $(1, D-1)$-dimensional 
QFT.  In string theory, the worldsheet is 
already two-dimensional,  
so we have a $(1,1)$-dimensional QFT theory already in the
first quantization.  

	First, let us recall briefly the results for point particles.  
Quantization replaces $P_\mu$ by $-i \frac \pa {\pa X^\mu}$, so the 
Einstein constraint on physical states (\ref {ptconstraint}) 
becomes the Klein-Gordon equation on wavefunctions.  
In a similar 
vein, the quantization of supersymmetric (spinning) particle would 
give rise to the Dirac equation as the constraint equation, as 
we will discuss in lecture three.

Now for strings, let us choose the conformal gauge (\ref {cfgauge}).  
In this gauge, the equation of motion (\ref {cfeom})
and the expression for canonical momentum (\ref {cfmom}) can 
be obtained from the action 
\beqar	
	S &=& \frac 1 {4 \pi \alpha'} 
			\myI {d^2\sigma} \sqrt {- \gamma} \gamma^{ab} 
				\pa_a X^\mu \pa_b X_\mu \nonumber \\
	  &=&  \frac 1 {\pi \alpha'} 
			\myI {d^2\sigma} \pa_+ X^\mu \pa_- X_\mu.
\label {cfact}
\eeqar
By varying the worldsheet metric away from $\gamma_{ab}$,
 we can find its (worldsheet) energy-momentum tensor 
$T^{ab} \sim \frac {\delta S} {\delta \gamma_{ab}}$.  
Since the action is conformally invariant, 
the trace of the classical energy-momentum tensor $T$ vanishes.  The remaining 
two components are
\beq	\label {streetensor}
	T_{++} = \frac 1 {\alpha^\prime} (\pa_+ X)^2, \sepe
	T_{--} = \frac 1 {\alpha^\prime} (\pa_- X)^2.
\eeq
The reparametrization invariance of the Nambu-Goto action implies
the first class constraints $T_{++}=0$ and $T_{--}=0$. In fact these
are the Virasoro constraints (\ref {constraint}) that we have seen before.

Canonical quantization  gives\myftnote 
	{In these lectures, $\imath$ denotes $\sqrt {-1}$.} 
\beq
	\comm {
\dot X^\mu(\sigma^0, \sigma^1)} {X^\nu(\sigma^0, {\sigma^1}')}
		= 2 \pi \alpha^\prime \imath \eta^{\mu \nu} 
			\delta ({\sigma^1}' - \sigma^1)
\eeq
The $X$'s must be periodic in $\sigma^1$ with period $2\pi$.  
After Fourier decomposition, we separate and recover the 
center of mass operators and the mode operators 
corresponding to excitations:
\[
	X^\mu = x^\mu + \alpha^\prime p^\mu \sigma^0 
			+ \sqrt{\frac {\alpha^\prime} 2} 
				\sum_{n\neq 0} \frac \imath n 
					\{\alpha_n^\mu e^{-\imath n \sigma^{+}}
					 +\tilde{\alpha}_n^\mu e^{-\imath n \sigma^{-}}\}
\]
\[ 
	[x^\mu, p^\nu] = \imath \eta^{\mu \nu}
\]
\beq	\label {xmode}
	[\alpha_n^\mu, \alpha_m^\nu] = n \eta^{\mu \nu} \delta_{n+m, 0}, 
~~~ [\tilde\alpha_n^\mu, \tilde\alpha_m^\nu]
= n \eta^{\mu \nu} \delta_{n+m, 0}
\eeq
\[
	(\alpha_n^\mu)^\dagger = \alpha_{-n}^\mu, \sepe
	(\tilde\alpha_n^\mu)^\dagger = \tilde\alpha_{-n}^\mu.
\]
So the Hilbert space is the tensor product of 
$2 \times D$ infinite towers of harmonic oscillators, 
each labeled by positive integers 
(coming from $\alpha_n$ and $\tilde\alpha_n$) 
and that of the $D$-dimensional 
quantum mechanics (coming from the zero modes $X^\mu$ and $P^\mu$):
\[	\bigotimes_{0 \leq \mu < D}^{n > 0} 
	\brac {\paren {\alpha_{-n}^{\mu}}^i \ket 0 
			| i = 0 \ldots \infty} 
	\bigotimes_{0 \leq \mu < D}^{n > 0} 
	\brac {\paren {\tilde \alpha_{-n}^{\mu}}^i \ket 0 
			| i = 0 \ldots \infty}
	\bigotimes \Phi (X^\mu).
\]
%
%
The operator $\alpha^\mu_{-n}$ ($\alpha^\mu_{n}$), with $n > 0$, 
creates (destroys) a quantum of left moving oscillation with angular 
frequency $n$ along the $X^\mu$ direction in the spacetime.  
$\tilde \alpha^\mu_{-n}$ ($\tilde\alpha^\mu_{n}$) 
does the same for the right movers.  This decomposition of degrees 
of freedom into essentially decoupled left and right movers is 
what makes many two-dimensional field theories so much more manageable 
compared to theories in higher dimensions.

	One should note that because of the indefinite signature of the spacetime 
metric $\eta$ in (\ref {xmode}), the states created by the oscillators 
along the time direction 
may have negative norms.  Such states are called \emph {ghosts}, not to be confused with 
the Faddeev-Popov ghosts, which also enter the scene later.  They cannot be present 
in the physical spectrum.  As we will see presently, the quantum mechanical 
implementation of the constraints (\ref {constraint}) eliminates them.

	It also is convenient to Fourier transform the energy-momentum
 tensor $T$:
\[
	T \equiv T_{++} = \frac 1 {\alpha^\prime} (\pa_+ X)^2 
		\equiv \sum_n L_n e^{- \imath n \sigma^+};	\]
\[	\tilde T \equiv T_{--} = \frac 1 {\alpha^\prime} (\pa_- X)^2 
		\equiv \sum_n \tilde L_n e^{\imath n \sigma^+};	\]
\[
	L_n = \sum_m \frac 1 2 \alpha_{n-m} \alpha_m, 
~~~ \tilde L_n = \sum_m \frac 1 2 \tilde \alpha_{n-m} \tilde \alpha_m, \]
\[ \alpha_0^\mu  = \tilde \alpha_0^\mu
=\sqrt{\frac {\alpha'} 2} p^\mu.
\]
These
$L_n$ and $\tilde L_n$ are well defined except for $n = 0$, for which there is 
a normal ordering ambiguity.  If we define
\beqar
	L_0 &=& \sum_{n>0} \alpha_{-n} \alpha_{n} + \frac 1 2 (\alpha_0)^2, 
				\nonumber \\
	\tilde L_0 &=& \sum_{n>0} \tilde \alpha_{-n} 
					\tilde \alpha_{n} + \frac 1 2
(\tilde \alpha_0)^2,
				\label {L0defn}
\eeqar
the constraint for the $n=0$ part would be $L_0-a = 0$, $\tilde{L}_0 -
\tilde{a} = 0 $ where $a$ and $\tilde{a}$ are constants reflecting
the normal ordering ambiguity. The combination $(L_0 + \tilde{L}_0)$ is
the Hamiltonian of the system generating a translation in $\sigma^0$
direction and $(L_0 - \tilde{L}_0)$ is the worldsheet momentum. Since
\[	[L_0, \alpha_{-n}] = n \alpha_{-n},	\]
the $n$-th oscillator has energy $n$, equal to its angular frequency.
The same holds for the right movers.

	We can try imposing 
\beq
	L_n - a \delta_{n, 0} = 0,~~
 \tilde L_n - \tilde a \delta_{n, 0}=0.
\eeq
for all $n$, as constraints on physics states.  
However we run into problems immediately.
It can be checked that 
the $L$'s form a representation of the \emph {Virasoro algebra}, which 
is the conformal algebra (\ref {cfalgebra}) with a nontrivial  
\emph {central extension}:
\beqar	\label {virasoro}
	\comm {L_n} {L_m} &=& (n-m) L_{n+m} 
			+ \frac c {12} (n^3 - n) \delta_{n+m, 0}, \nono
	\comm {\tL_n} {\tL_m} &=& (n-m) \tL_{n+m} 
			+ \frac c {12} (n^3 - n) \delta_{n+m, 0}, \nono
	\comm {L_n} {\tL_m} &=& 0.	
\eeqar
In our case, the central charge $c$ is equal to $D$, the  
spacetime dimension.
Imposing $L_n |phys \rangle = 0$ for all 
$n \in \intZ$ would be inconsistent with the commutation
relation if $D \neq 0$.   We may instead adopt the
Gupta-Bleuler prescription and require that physical states 
be annihilated by half of $L_n$'s
\beq \label {phystatecond}
	(L_n - a \delta_{n,o}) |phys \rangle
 = 0, \come n \geq 0.
\eeq
We also define an equivalence relation among them: 
\beq \label {strgaugetran}
	\ket {phys} \sim 
   \ket {phys} + L_{-n} \ket {*}, 
		\come n > 0.
\eeq
We call a physical state \emph {spurious} if it is a linear 
combination of $L_{-n} \ket {*}$ for some state
$\ket{*}$.  The true physical 
degrees of freedom are thus the equivalence classes of (\ref {strgaugetran}).
Condition (\ref {phystatecond}) implies that the matrix elements 
of $L_n$ between physical states vanish for all $n$.  
This is consistent with 
(\ref {strgaugetran}), which says that any $L_n$ has no physical 
effect on a physical state.  It is the same story for 
the right movers.

As alluded earlier, these $2 \times \infty$ set of constraints and 
equivalence conditions effectively remove $2$ directions of 
oscillators of every mode from the physical spectrum 
if the theory is consistent.  In the 
next section we demonstrate this explicitly for the 
first few excited 
states.  Since $L_n$ and $\tilde L_n$ have exactly 
the same property, in the following discussion we will concentrate 
on $L_n$, bearing in mind that the same results obtain for $\tilde L_n$.
In particular, we will determine $a$ by 
consistency requirement. Since we can repeat the same story for $\tilde a$,
they must be the same, implying
\beq \label {levelmatch}
	(L_0 - \tilde L_0) \ket {phys} = 0.
\eeq
This is known as the {\em level matching condition}.

\subsection {Critical Dimensions}

Another way to quantize the string is to start with the light-cone 
action and perform canonical quantization. In this gauge,  the constraint equations 
(\ref {ptconstraint}) are explicitly solved and  only 
the $(D-2)$  oscillatory excitations transverse to  the string 
worldsheet remain.  Whether we choose the light-cone gauge or
the conformal gauge is a matter of convention and their results should
agree unless   there is an anomaly obstructing
conformal invariance from becoming a full fledged quantum symmetry.  As 
conformal invariance is the remnant of gauge symmetry on the worldsheet, 
an anomaly for it would spell disaster. 

Let us look at the spectrum in the conformal gauge, taking into 
account the physical state condition (\ref {phystatecond}) 
and the string gauge covariance (\ref {strgaugetran}).  
As a measure of oscillator excitation, 
define
\beq	\label {bosonstrength}
	N \equiv L_0 - \frac 1 2 (\alpha_0)^2 
		= \sum_{n>0} n \alpha_{-n} \alpha_{n}
\eeq
and similarly $\tilde N$ for the right moving sector.  By 
(\ref {L0defn}) and the Einstein relation $m^2 = - k^2$, 
they also determine the mass of the states: 
\beq	\label {stringmass}
	m^2 = \frac 4 {\alpha^\prime} (N - a).
\eeq
Therefore the constraint $L_0 = a$ is the \emph {mass shell condition}.
The level matching condition (\ref {levelmatch}) implies 
that $N = \tilde N$.  As mentioned 
above, it is sufficient to concentrate on the left movers.

\paragraph {Ground State --- $N=0$.}
	There is no oscillator excitation and the states are simply 
$|k \rangle$ where
\[ p^\mu \ket k  = k^\mu \ket k , ~~
   \alpha_n \ket k  = 0 ~~(n>0) . \]
The only nontrivial condition from (\ref{phystatecond}) is the mass
shell condition
\[
	(L_0 -a ) |k\rangle = (\frac {\alpha^\prime} 4 k^2 - a) |k\rangle  = 0,
\]
which implies 
\[		m^2 = - \frac 4 {\alpha^\prime} a. \]
If $a > 0$, then the ground state would correspond to a tachyon.  As it turns out, 
this is indeed the case for both bosonic string and superstring theory.  In the 
latter, we will be able to consistently truncate the spectrum
of the superstring  so that the 
ground state tachyon is no longer present, but this seems impossible 
for the bosonic string theory.  As we know from field theory, the presence 
of a tachyon indicates that we are perturbing around a local maximum of 
potential energy --- we are at a wrong vacuum.  However, to this date it 
is not known whether one can find an alternative vacuum 
for the the bosonic string theory so that it is free 
of tachyons.

\paragraph {First excited level --- $N = 1$.}
 The states are 
$e_\mu(k) \alpha_{-1}^\mu \ket{k}$.
It is simple to deduce from (\ref {phystatecond}) the 
following constraints:
\beqarn
	(L_0 - a) \ket{phys} = 0 &\rightarrow& 
		k^2 = (a - 1)  \frac 4 {\alpha^\prime}, \\
	L_1 \ket{phys} = 0 &\rightarrow& k \cdot e = 0.
\eeqarn
The equivalence relation (\ref{strgaugetran}) states that 
\beq
	e_\mu \sim e_\mu + \lambda k_\mu,
\eeq
which has precisely the form of a gauge transformation in 
QED.  However, this does not yields a spurious physical 
state unless $k^2 = 0$.  This is 
familiar from QED: if states on this level are massless 
(i.e. $a = 1$), then physically there are only $(D-2)$ 
independent polarizations; otherwise 
there are $(D-1)$ polarizations.  Since the light-cone 
gauge quantization gives $(D-2)$ polarizations, 
the anomaly-free requirement picks $a = 1$.
Incidentally, this result can also be obtained if we 
determine the normal
ordering prescription of $L_0$ in light-cone gauge by the 
$\zeta$-function regularization.

	Also by analogy to QED, $k \cdot e = 0$ can be interpreted as the 
Lorentz gauge condition.  Combined with $k^2 = 0$, the massless 
Klein-Gordon equation, we obtain the Maxwell equation $\partial^\mu
F_{\mu\nu} = 0$.  These statements are 
precise in open string theory.  In closed 
string, when we combine them with the their counterparts for the right 
movers, we obtain the equations of motion and gauge transformations 
appropriate for graviton, antisymmetric tensor, and dilaton fields.

\paragraph {2nd excited level --- $N = 2$.}
The states take the form
	$[e_{\mu \nu} \alpha_{-1}^\mu \alpha_{-1}^\mu  + e_\mu \alpha_{-2}^\mu]
		\ket {k}$.  
The mass shell condition reads
\[ (L_0 -1) = 0 \rightarrow k^2= - \frac 4 {\alpha^\prime}, \]
where we have used the result $a = 1$.  
This means states at this level 
are massive.  The other two nontrivial 
physical state conditions from $L_2$ and $L_1$ 
impose $(D+1)$ conditions on $e_{\mu \nu}$ and $e_\mu$.  
They leave us 
with 
\[	\frac 1 2 D(D+1) + D - (D+1) = \frac 1 2 (D^2 + D) - 1	\]
degrees of freedom in the polarization.  On the other hand, light-cone 
quantization gives 
\[	\frac 1 2 (D-1)(D-2) + D-2 = \frac 1 2 (D^2 - D) - 1	\]
degrees of freedom.  The deficit of $D$ must be accounted for 
in the equivalence relation (\ref{strgaugetran}).  The spurious 
states at level two are spanned by \myftnote
		{The particular choice of basis given here is 
		purely a matter of convenience.}
\[	\chi_\mu L_{-1} \alpha_{-1}^\mu \ket {k}	\]
and 
\[	(L_{-2} + \gamma L_{-1}^2) \ket{k},	\]
for some constants $\chi_\mu$ and $\gamma$.
Requiring the first type of spurious states to be physical 
leads to the condition
\[	\chi \cdot k = 0. \]  
Therefore the spurious state of the first type accounts for $(D-1)$ degrees of 
freedom.  Since we need to have $D$ spurious states, 
the second type of state must also satisfy the physical state condition.
It is left to students to verify that the $L_1$ constraint
requires $\gamma$ to be $\frac 3 2$ and the $L_2$ constraint fixes $D$ to
be 26.
Thus we have arrived at the famous conclusion that bosonic string 
theory propagates in 26 dimensions.

	We can continue this program to states of higher levels.  The result 
is the same --- only for $a=1$ and $D=26$ do we have an agreement between 
light-cone and conformal gauge.  If one insists on considering $D\neq26$
nonetheless, then it has been found that spacetime Lorentz invariance is 
completely lost in the light-cone gauge 
unless $D=26$ (ref. 204 in \cite{GSW}, Vol 1).
On the side of the conformal gauge, 
although one can show there is no ghost in the tree level 
spectrum if $D\leq 26$
(refs. 65 and 202 of \cite {GSW}, Vol 1), 
they do show up as unphysical poles in one-loop string amplitudes unless 
$D=26$.  Below we will mention very briefly the correct 
formulation 
of such \emph {non-critical} 
string theory found by Polyakov (ref. 366 in \cite{GSW}, Vol 1).  

\subsection {Massless Spectrum}

	Now let us examine the spectrum of massless 
states in 26-dimensional bosonic string theory.  
According to (\ref {stringmass}), 
the masses of the string states 
are integral multiple of $2/{\sqrt {\alpha'}}$.   
 From last section, we see the massless particles arise from the first
excited level of the string.
Combining left and right moving sectors of the Fock space in accordance with 
the level matching condition (\ref{levelmatch}), they 
have the form 
\beq	\label {nomstate}
	e_{i j} \, \alpha_{-1}^i \tilde \alpha_{-1}^j \ket{k},~~ 
		(k^2=0; i, j= 1, \ldots 24)
\eeq
in the light-cone gauge. 
We may decompose $e_{ij}$ into irreducible representations 
of $SO(24)$, each of which would correspond to a certain type of particle:
\beqarn	e_{i j} 
 &=& [\frac 1 2 (e_{i j} + e_{j i}) 
					- \frac 1 {12} \delta_{i j} \Tr e ]
			  	+ [\frac 1 2 (e_{i j} - e_{j i})] 
			  		+ [\frac{1}{24} \delta_{i j} \Tr e] \\
			&\equiv& [h_{i j}] + [B_{i j}] + [\delta_{i j} \Phi] .		  
\eeqarn
The traceless symmetric, antisymmetric, and trace parts of $e_{ij}$
are denoted as $h_{i j}$, $B_{i j}$ and $\Phi$ respectively. 
$B_{i j}$ is known as the 
\emph {antisymmetric tensor}.
 $\Phi$ is called \emph {dilaton}.  
Being a massless scalar, $\Phi$  may develop
a vacuum expectation value (VEV).  
We will later see that $\langle \Phi \rangle = \Phi_0$ shifts 
the string coupling constant $\kappa$  to $\kappa e^{\Phi_0}$.  
$h_{i j}$ is identified with the
graviton because it observes general 
covariance.  To see this we should choose the conformal gauge, 
which is $26$-dimensional Lorentz covariant.  Now we use 
Greek indices 
$\mu$, $\nu$, \ldots, ranging from 0 to 25, to label 
the tangent space.  We mentioned earlier 
that the equivalence relations from 
$L_{-1}$, $\tilde L_{-1}$ have the spacetime interpretation of 
gauge transformations.  It is not difficult to show that
these gauge transformations act on $h_{\mu\nu}$ and $B_{\mu\nu}$
as
\beqarn
	h_{\mu \nu} &\rightarrow& h_{\mu \nu} + \pa_\mu \xi_\nu + \pa_\nu \xi_\mu, \\
	B_{\mu \nu} &\rightarrow& B_{\mu \nu} + \pa_\mu \xi_\nu - \pa_\nu \xi_\mu.
\eeqarn
The first is simply diffeomorphisms acting on the spacetime metric in 
the Minkowski background.  The second can be written in the 
language of differential forms as 
$B \to B + dA$.  This suggests that the physical 
observable associated 
with the 2-form $B$ should be 
its \emph {3-form field strength} $H \equiv dB$.

\subsection {Polyakov Action}

There is another interpretation of the requirement $D=26$, 
due to Polyakov. Consider the action 
\beq	\label {polyakovact}
	S = \frac 1 {4 \pi \alpha^\prime} 
			\int \! d^2\sigma \, \sqrt {- g} g^{a b} \pa_a
X^\mu  \pa_b X_\mu,
\eeq
where both, the metric $g_{ab}$ as well as $X^\mu$, are
treated as dynamical variables.  The worldsheet
metric $g_{ab}$ has no local propagating degrees of freedom. 
Classically, the equation for $g$ 
requires it to be proportional to the induced metric 
(\ref{indmet}).  Substitute this back to 
(\ref {polyakovact}) and we obtain the Nambu-Goto action 
(\ref{nambugoto}), establishing their 
classical equivalence.

In fact the worldsheet metric
consists almost purely of gauge degrees of freedom.  First 
the worldsheet metric has three independent degrees of freedom, 
two of which can be gauged away using worldsheet diffeomorphism, 
bringing the metric into the standard form
\beq		\label {cfgauge2}
	\paren {g_{ab}} = \left( \bear {r r}
						- \lambda & 0 \\
				  		   0	  & \lambda
				  	\ear \right),
\eeq
in what is known as conformal coordinates.  Furthermore,
the Polyakov action (\ref {polyakovact}) has the 
Weyl rescaling symmetry which allows us to scale $\lambda$ to, say, $1$.  
In this gauge, the equation of motion and canonical momenta 
can be obtained from the same conformal gauge action as for the Nambu-Goto 
action (\ref {cfact}), so the same quantization procedure can be carried 
over.

There are two complications to this story.  First, in general 
(\ref {cfgauge2}) can only be enforced in each coordinate patch.  
Between patches there can be global degrees of freedom left.  Roughly 
speaking they describe the shape of the worldsheet and are known 
as \emph {complex moduli}, for reasons to be discussed in Greene's
lecture at 
this school.  A simple example appears in the next section.  
Second, quantum mechanically the Weyl rescaling symmetry 
may became anomalous, and the algebra of conformal transformation 
(\ref {cfalgebra}) is not realized on the Hilbert space.  
It is deformed to be the Virasoro algebra with the central
extension.  
The central charge $c$ measures the violation of conformal invariance.
As we saw, the central charge for $X^\mu$ is $D$, equal to the
dimension of the spacetime.   
The Faddeev-Popov ghosts,  which provide the correct 
normalization for the path integral
respecting the reparametrization invariance,
carry central charge $-26$. Since the conformal anomaly is additive,
only when $D=26$ does the anomaly from the $X$'s cancel 
against that from the ghosts and give us a consistent 
theory.  When $D \neq 26$, one can no longer gauge away 
$\lambda$ and has to treat it as dynamical  
degrees of freedom, known as \emph {Liouville field}.  
The resulting theories, known as \emph {non-critical 
string}, are interesting in their own right but will not be 
discussed in this lecture. For reviews on this topic, 
see \cite{gm} and \cite{joe}. 

\subsection {String Propagation and Interactions}

\myepsf {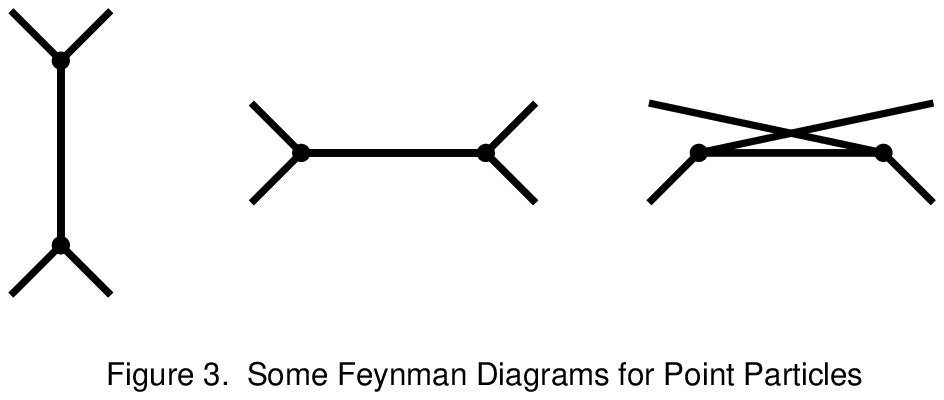}
\myepsf {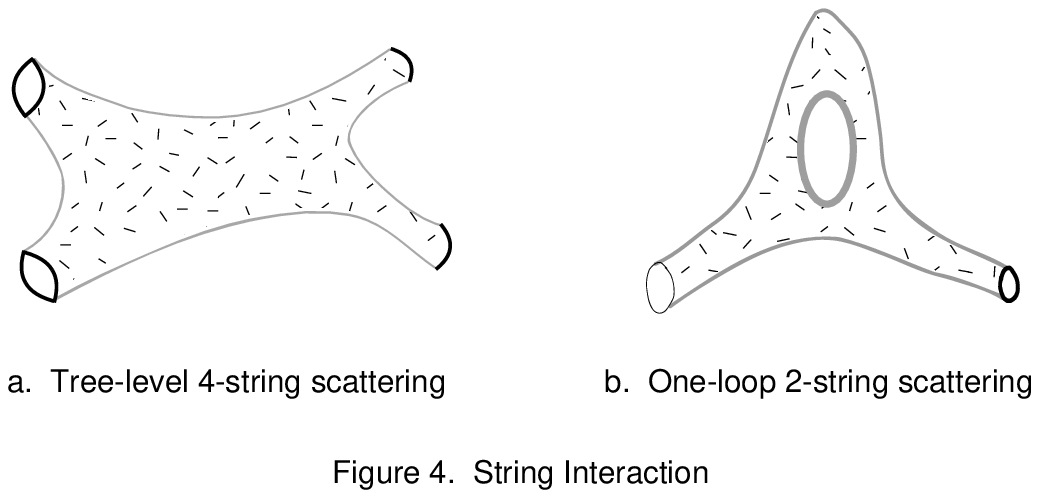}
	
	Point particles propagate in a straight line with amplitude 
given by their Feynman propagators.  They interact at a well-defined 
point in the spacetime, where straight lines intersect at vertices. 
Each vertex also has some coupling constant associated with it.  We 
calculate a scattering amplitude of them by drawing 
the corresponding Feynman diagrams, and multiplying 
together all the propagators  and 
the coupling constants at each vertex  (fig. 3).  
In string theory, the picture is similar (fig. 4).  
Propagation of string is represented by a tube.  A slice of
the worldsheet at any time determines a
string state at that instant.  
However, because of worldsheet reparametrization invariance, no 
scheme of time slicing is preferred over others.  This and the 
smooth joining and splitting of string tubes mean that there is no 
freedom in assigning 
coupling constants to any particular point.  Indeed it will soon 
become clear that there is only one measure of string coupling, 
which is however a field carried by and distributed over the strings 
themselves.

To study string worldsheets of various topologies, it is convenient
to choose the worldsheet metric to be Euclidean rather than Lorentzian.  
This can be done by performing a Wick rotation on the worldsheet: 
\[	\sigma^0 = - \imath \sigma^2	\]
\[	z \equiv \imath \sigma^+  = \sigma^2 + \imath \sigma^1, \come 
	\zb = \imath \sigma^-  = \sigma^2 - \imath \sigma^1,	\]
\[	X^\mu = x^\mu - \imath \alpha' p^\mu \re z
			+ \sqrt {\frac {\alpha'} 2} 
			  \sum_{n \neq 0} \frac \imath n 
			  \{ \alpha_n^\mu e^{- n z} 
			  + {\tilde {\alpha}}_n^\mu e^{- n \zb} \}	\]
We will use this Euclidean notation from now on.

	Figure 4 shows the worldsheet for a string-string 
scattering.  Its amplitude is calculated by evaluating the Polyakov 
path integral over it.
After gauging away arbitrary reparameterizations, 
the integration over the worldsheet metric $g$ of Polyakov 
action is reduced to a sum of over all possible shapes and 
sizes of worldsheets of a given topology.  
Since the size of the worldsheet can be gauged away for critical 
string theory,  this reduces to a finite 
dimensional integral over its moduli space, the space that 
parameterizes the shape of worldsheet with this topology.  
Worldsheet actions themselves do not tell us 
which topology of worldsheet we should choose, but 
analogy with Feynman diagrams suggests 
that handles in the worldsheet represent 
internal loops and we should sum over all 
number of handles.  In fact the unitarity of the $S$-matrix
dictates how to sum over topologies of the worldsheet.  

	As a simple and useful illustration, 
consider the one-loop vacuum to vacuum string amplitude (fig. 5).  
This has the physical interpretation of calculating 
the vacuum energy.  There 
is no external string and the worldsheet is topologically 
a torus.  By Weyl scaling we can always make it a flat 
torus, defined as the quotient of the complex plane by a 
lattice generated by $1$ and $\tau$ --- we  
identify points related by $n + m \tau$, 
$n, m \in \intZ$ (fig. 6).  
$\tau$ is the complex moduli for the topological class of 
torus and cannot be gauged away by Weyl rescaling.  
The integration over $g$ now reduces to an integration 
over the moduli parameter $\tau$\myftnote
	{However there are further discrete identification 
	due to large diffeomorphisms, to be discussed in 
	lecture two.}. 
Nondegeneracy 
of the torus requires $\im \tau \neq 0$, and by choices of 
basis of lattice vector we can require $\tau$ to live on 
the upper complex half-plane.  Let us look at this from the worldsheet 
viewpoint.  Choose the imaginary axis as worldsheet ``time'' 
and real axis as the spatial extent of the string.
Then $\im \tau$ is the worldsheet time.  Worldsheet states evolve 
along it as usual with Hamiltonian $L_0 + \tL_0$.  
$\re \tau$ is a spatial twist, generated by the 
worldsheet momentum $L_0 - \tL_0$.  As there is no end to 
the string in this one-loop amplitude, the path integral 
sums over all states in the Hilbert space --- it is 
a trace.  In fact it is the partition 
function
\beqar	\label {parti}
	Z(q) &\equiv& \Tr \: q^{L_0 - \frac {D-2} {24}} 
			\bar q^{L_0 - \frac {D-2} {24}} \nono
	&=& (2 \im \tau)^{-D/2} (q \bar q)^{- \frac {D-2} {24}}
		\abs {\prod _{n=1}^{\infty} \frac 1 {(1-q^n)^{D-2}}}^2, \nono  
	&& q = e^{2\pi \imath \tau},\come \im \tau \geq 0.
\eeqar
The number $\frac 1 {24}$ in the exponent is due to the conformal 
anomaly.  A proper explanation 
would take us too far afield, but it can be found in, for example, \S 7.1 of 
\cite {ginsparg}.  The combination $(D-2)$ is easy to 
understand in light-cone gauge, but can also be obtained in 
the conformal gauge if one also includes the contribution from 
the Faddeev-Popov ghosts.  
Here it is sufficient to note here that 
with $D = 26$ we recover the correct 
normal ordering constant $a = \tilde a = 1$.    We also note that the 
mass for states in a level can be read from the corresponding 
exponent for $q$ and $\bar q$ outside the $(2 \im \tau)^{-D/2}$ 
factor. For example, the exponent for the tachyon 
is negative, and that for massless states are zero.
The  $(2 \im \tau)^{-D/2}$ factor is the result 
of momentum integration, and here we have
$D$ rather than $(D-2)$ since the string zero modes are not affected by the
light-cone gauge condition.   
The coefficient in front of a monomial in $q$
counts the multiplicity of states with the corresponding mass
($=$ the degree of the monomial).  
For example, from (\ref {parti}) one sees that there is just one
tachyon. To complete the 
calculation of the amplitude, one also needs to integrate over the 
moduli parameter $\tau$, which parameterizes the length 
and twist of the torus as discussed earlier.  Observe 
that the integration over $\re \tau$ enforces the 
level matching condition, as those terms with unequal exponents for 
$q$ and $\bar q$ will vanish.
  
\myepsf {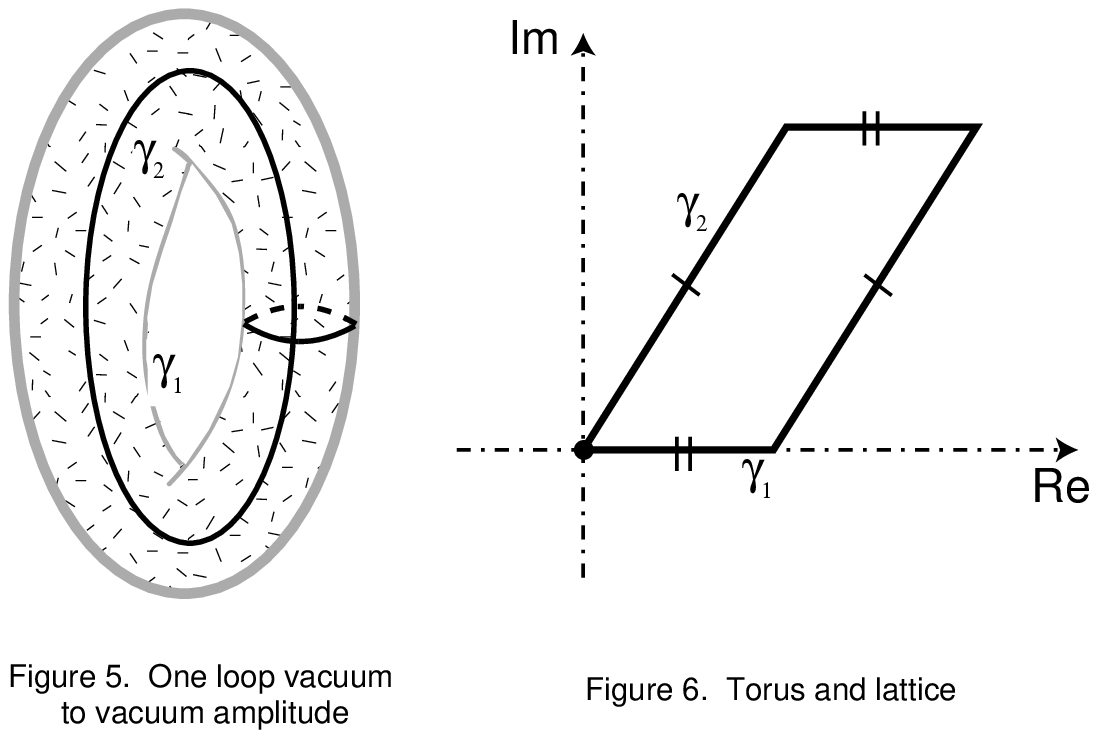}

\subsection	{Conformal Field Theory}

The conformal gauge action (\ref {cfact})
 is an example of a 2d \emph {conformal field theory} (CFT).
Although the details of CFT are outside 
the scope of this lecture (for extensive discussion on the subject,
see for example \cite{ginsparg} and \cite{lust}), 
we will now introduce some facts 
and concepts that  will be  useful.  Consider a path 
integral calculation of a CFT over some Riemann surface, with 
some tubes extending to infinity.  The field configurations 
at the ends of the tube correspond to states in the CFT Hilbert 
space.  In string theory they represent external, asymptotic 
string states in a scattering process.
We can perform arbitrary conformal transformations 
when evaluating the path integral of a CFT.  Let us 
choose one that brings the tube C in (fig. 7) from 
infinity to within a finite distance from the scattering 
region.  Because this would involve an infinite 
rescaling in the neighborhood of the end circle of tube C, 
the end circle, which has finite radius, will shrink to a point.  
Its effect should therefore be represented by the insertion of 
a local field operator at that point.  It is called a \emph {vertex operator}.  
Therefore there is a one-to-one correspondence between states and 
operators in CFT.  In string 
theory, for example, a vertex operator taking momentum $k$ 
has the form, : (oscillator part) $\times {e^{ik\cdot X}}$ :,
where :: denotes the normal ordering.  
The oscillator part of the operator is determined by its counterpart 
for the corresponding state.  For example, the operator 
that creates an insertion of a massless operator of momentum $k$ is
\[	\xi_{\mu\nu}\nod{ \pa X^{\mu} \pb X^\nu e^{ik\cdot X}}\,.	\]
For the tachyon, the oscillator part is just the
 identity, so the vertex operator 
is simply $:e^{ik\cdot X}:$.
Of course, not all vertex operators correspond to insertion of physical 
states.  They have to obey the operatorial version of 
the physical state condition  
(\ref {phystatecond}).  The condition for the tachyon is simply 
\(k^2 =  4 / {\alpha'}\).  

\myepsf {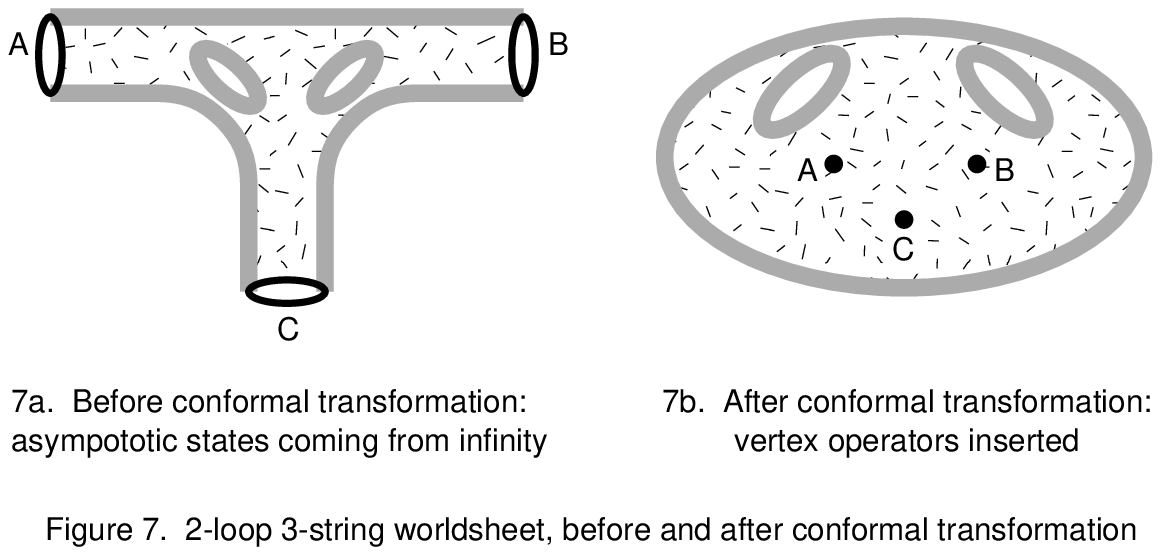}

	States that satisfy (\ref {phystatecond}) 
with $a$ and $\tilde a$ not necessarily equal to $1$ are called 
\emph {Virasoro primary}
states  of \emph {conformal weight} ($a$, $\tilde a$)\myftnote
	{So physical states are Virasoro primary states of conformal 
	weight $(1,1)$}.  
The corresponding operators are called \emph {Virasoro primary fields}.  
For a Virasoro primary operator $\Phi$, its defining properties  
can be summarized in the 
\emph {singular parts} 
of its operator product expansion (OPE) with the energy-momentum tensor:
\beqar	\label {virasoroope1}
		T (z) \phi (w, \bar w) &\sim &
			\frac {a \phi (w, \bar w)} {(z-w)^2} 
			+ \frac {\pa \phi(w, \bar w)} {(z-w)},	\nono
		\tilde T (\bar z) \phi (w, \bar w) &\sim& 
			\frac {\tilde a \phi(w, \bar w)} {(\bar z - \bar w)^2} 
			+ \frac {\pb \phi(w, \bar w)} {(\bar z - \bar w)}.
\eeqar
The Virasoro algebra (\ref {virasoro}) itself can be written as 
\beq	\label {virasoroope2}
		T(z) T(w) \sim \frac {c/2} {(z-w)^4} 
					+\frac {2 T (w)} {(z-w)^2} 
			+ \frac {\pa T(w)} {(z-w)},
\eeq
and similarly for $\tilde T$ with no singularity between $T$ and $\tilde T$.
Thus $T$ is almost a Virasoro primary field of weight $(2, 0)$ except for 
its conformal anomaly.
It is a fundamental property of a conformal 
field theory that its 
Hilbert space and operator content is a direct sum of often 
infinitely many irreducible 
representations of the left $\times$ right Virasoro algebra, each of which 
is generated by the action of the algebra on a highest weight 
state.  The Virasoro primary fields of a CFT and their operator 
product expansion (OPE) completely characterize it.
For later use, let us state the OPE between 
basic fields in the bosonic action (\ref {cfact}):
\beqar	\label {usefulope}
	\pa X^\mu (z) X^\nu (w) &\sim& 
		\frac {- \frac {\alpha'} 2 \eta^{\mu\nu}} {(z-w)}; \nono
	\pa X^\mu(z) \nod {e^{\imath k \cdot X(w)}} 
		&\sim& \frac {- \imath \frac {\alpha'} 2k^\mu} {(z-w)}; \nono
	\nod {e^{\imath k_1 \cdot X(z)}} \nod {e^{\imath k_2 \cdot X(w)}}
		&\sim& \abs {z-w}^{\alpha ' k_1 \cdot k_2} 
			\nod {e^{\imath k_1 \cdot X(z) + \imath k \cdot X(w)}}.
\eeqar

\subsection {Low Energy Effective Action}

	Let us sum over momentum and make a Fourier transformation, 
then the vertex operators for the massless particles are
\[h_{\mu\nu}(X(z,\bar z)) \partial_z X^\mu \bar{\partial}_{\bar{z}}
 X^\nu \]
for the graviton field and  
\[ B_{\mu\nu}(X(z,\bar z)) \partial_z X^\mu \bar{\partial}_{\bar{z}}
 X^\nu \] for the antisymmetric tensor field.  
Now consider inserting coherent states of these fields --- 
exponential of their integral over the worldsheet --- in every 
correlation function we compute for the Polyakov action (\ref {polyakovact}).  
Physically, this should be interpreted as vacuum expectation values
for these spacetime fields.  From the worldsheet viewpoint, they 
simply modify the action (\ref {polyakovact}) into 
\beq \label {bgone}
	S = \frac 1 {4 \pi \alpha^\prime}
			\myI {d^2\sigma} 
			  \{(\sqrt g g^{ab} G_{\mu\nu}(X) + \epsilon^{ab} B_{\mu\nu}(X))
			  		\pa_aX^\mu \pa_bX^ \nu \},
\eeq
where $G_{\mu\nu} = \eta_{\mu\nu} + h_{\mu\nu}$.
\noindent
We can also introduce a super-renormalizable term to (\ref {bgone}) 
which corresponds to a VEV for the tachyon.  
Noting that the worldsheet scalar fields such as $X^\mu$ have zero scaling 
dimension, it is easy to see that the result is the most general 
renormalizable action we can write with $X^\mu$ and their derivatives in two
dimensions.  However, if we add a background for any one of the massive 
states, the corresponding operator would be non-renormalizable and 
would in general generate terms corresponding to the VEV's for all other 
massive states.  

Students may notice that the dilaton $\Phi$ is missing in this
discussion. If we allow ourselves to use the worldsheet metric $g_{ab}$
in addition to the scalar field $X^\mu$, there is another operator of
dimension two on the worldsheet, $R\Phi(X)$, where $R$ is the worldsheet
curvature. It is a long story to explain why this is a proper coupling
of the worldsheet to the dilaton field\myftnote
	{Very briefly, this coupling is obtained by regularizing the 
	dilaton vertex operator on a curved worldsheet and rescaling 
	the background metric.}.  
The complete worldsheet action under
the background of the massless fields is then 
\beq		\label	{bgstract}
S = \frac 1 {4 \pi \alpha^\prime}
\myI {d^2\sigma} 
\{(\sqrt g g^{ab} G_{\mu\nu}(X) + \epsilon^{ab} B_{\mu\nu}(X))
	\pa_aX^\mu \pa_bX^ \nu + \alpha' \sqrt{g} R \Phi(X). \}
\eeq
We note that if we let $\Phi \to \Phi + \Phi_0$, where $\Phi_0$ 
is a constant, then $S \to S + \Phi_0 \chi$, where $\chi=2-2h-b$ is the 
Euler number of the worldsheet surface, $h$
 is the number of handles and $b$ that of 
boundaries on the worldsheet. 
Since $h$ is the number of loops in the string ``diagram,'' 
shifting the dilaton 
field by a constant $\Phi_0$ is equivalent to multiplying the string  
loop expansion parameter $\kappa^2$  by $e^{2\Phi_0}$.  Looking  
closely enough, all string diagrams can be
seen as combinations of $\phi^3$ type of vertices 
and $\kappa$ their coupling constant.

	Recall that earlier on when we considered the simple case in which  
all the VEVs of these massless spacetime fields vanish, i.e. when 
the sigma model string action (\ref {bgstract}) is free, the decoupling 
of the conformal factor $\lambda$ in the metric $g_{ab}$ requires 
conformal invariance at the quantum level. This then led to the 
requirement of $D=26$.  Now with some of the VEV's being nonzero, 
quantum conformal 
invariance requires the vanishing of $\beta$ functions:
\beqar
	0 = \beta_{\mu\nu}^G 
	  &=&  \alpha^\prime R_{\mu\nu} 
	 	+ 2 \alpha^\prime \na_\mu \na_\nu \Phi
	 	- \frac {\alpha^\prime} 4  H_{\mu\lambda\rho} {H_\nu}^{\lambda\rho}
	 	+ O(\alpha^{\prime 2})	 \label {betag}		\\
 	0 = \beta_{\mu\nu}^B 
	  &=& - \frac {\alpha^\prime} 2 \na^\lambda H_{\lambda\mu\nu} 
	 	+ \alpha^\prime (\na^\lambda \Phi) H_{\lambda \mu\nu} 
	 	+ O(\alpha^{\prime 2})	\label {betab}		\\
	0 = \beta^\Phi 
	 & = & - \frac {\alpha^\prime} 2 \na^2 \Phi 
	 	+ \alpha^\prime (\na\Phi)^2 
	 	- \frac 1 {24} \alpha^\prime H^2 
	 	+ O({\alpha'}^{2})			\label {betaphi}
\eeqar
These can be regarded as the equation of motion coming from 
a spacetime action of $G$, $B$, and $\Phi$:
\beq	\label {sugraactstrmet}
	S = \frac 1 {2 \kappa^2} 
		\myI {d^{26}X} 
			\sqrt {-G} e^{-2\Phi} 
			\{ R - \frac 1 {12} H^2 + 4 (\na \Phi)^2 + O({\alpha^\prime}) \}.
\eeq
Here we see explicitly that the shift $\Phi \to \Phi + \Phi_0$ can be compensated by 
$\kappa \to \kappa e^{\Phi_0}$ for constant $\Phi_0$.

In this action, the normalization of the Einstein-Hilbert term is 
not standard, and the sign of the dilaton kinetic term is wrong.  
We can cure these problems by a field redefinition:
\beq	\label {str2einsteinmet}
	{G}_{\mu\nu} = e^{ - \Phi / 6} \tilde G_{\mu\nu},
\eeq
and the action 
(\ref {sugraactstrmet}) can be rewritten as 
\beq	\label {sugraeinsteinmet}
	S = \frac 1 {2 \kappa^2} 
		\myI {d^{26}X} 
			\sqrt {-\tilde G} 
			\{ \tilde R - \frac 1 {12} e^{- \frac 1 3 \Phi} H^2 
				- \frac 1 6 (\na \Phi)^2 + O({\alpha^\prime}) \}.
\eeq
Different choices of the metric correspond to different units of 
length (different rulers).
$\tilde G$ is known as the \emph {Einstein metric} while $G$ is called
the \emph {string metric}.

\section {Lecture Two: Toroidal Compactifications}

	The restriction on spacetime dimension by requiring quantum 
mechanical consistency is a striking result.  Some analog of it may 
one day tell us why we live in three spatial and one temporal dimensions.  
However, as a candidate theory of everything, string theory faces 
the immediate criticism that it gives us {\em too many} 
dimensions.  Later, when we come to superstring, the critical 
dimension will be lowered to $(9+1)$, but that is still $6$ dimensions 
in excess.  Naturally one entertains the possibility that the 
true spacetime takes the form of a direct product $M^4 \times K$, where 
$M^4$ is the $4$-dimensional Minkowski space we 
recognize everyday and $K$ an extremely 
tiny compact manifold that  our crude probes of nature have so far 
failed to reveal.  As you all know, this idea has existed in field theory 
in the form of Kaluza-Klein program long before string theory was 
invented.  However, as we will presently see, string compactifications 
introduce interesting ``stringy'' effects not seen in the usual 
Kaluza-Klein schemes.

	For a string propagating in a $M\times K$ background spacetime with 
constant VEV $\Phi$ for the dilaton, we may 
absorb $\Phi$ into the string coupling 
constant.  The conformal gauge action is then 
\beq	\label {cfactorus}
	S = \frac 1 {4 \pi} \myI {d^2z} (G_{\mu\nu} + B_{\mu\nu}) 
								\pa X^\mu \bar \pa X^\nu,
\eeq
where we have set $\alpha'$ to $2$ by choosing a unit of length.
Because of the direct product structure of $M\times K$, $S$ can be 
split into an external part $S_M$ involving 
coordinates on $M$ and an internal part $S_K$ on $K$, 
which can be studied separately.  The analysis of $S_M$ is trivial and 
all the interesting consequences of compactification come from $S_K$.
In this lecture we concentrate on the simplest possible choices for $K$:
the tori.  They are simply products of $S^1$ and are flat.  One can 
choose constant metrics for them and the nonlinear sigma models describing 
string propagating on them are still free as a two-dimensional QFT.  
We will consider the spacetime as being $M^{26-D} \times T^D$.  Although 
the ultimate goal of string theory is to describe the $D=4$ world we live in,
it turns out to be very instructive and enlightening to consider diverse 
choices of $D$.  We will encounter many ideas useful for the rest of these 
lectures as well as many others to come in this school.

\subsection {Lattice and Torus}

We can always parameterize a flat torus $T^D$ so that its metric $G_{ij}$ 
is constant and the coordinates $x^i$ have period $2\pi$, i.e. 
\beq	\label {torusdef1}
	T^D \equiv  {{\Bbb R}^D \over \sim}  ~,
\eeq
where
$$
	X^i \sim X^i + 2\pi m^i~,~~~~ m^i \in \intZ~. 
$$
We will use indices $i$,$j$,\ldots in this coordinate system.  
It turns out 
to be convenient to introduce a constant vielbein $e^a_i$ and 
new coordinates $X^a$ to bring 
the metric into the standard Euclidean form:
\beqarn	
	G_{ij} &=& e^a_i e^b_j \delta_{ab}, \sepe a = 1, \ldots, D.	\\
	X^a &\equiv& e^a_i X^i
\eeqarn
We use indices $a$, $b$, \ldots in these coordinates. In the
new coordinates $X^a$,   
the periodicity condition is changed to
\beq	\label {torusequiv}
	X^a \sim X^a + 2 \pi e^a_i m^i .
\eeq
In this way, instead of characterizing the size and shape of a torus 
by defining it with a fixed lattice ($\intZ^D$) as in (\ref {torusdef1}) 
with an arbitrary constant Riemannian metric $G_{ij}$,  we can use the 
fixed metric $\delta_{ab}$ and an arbitrary nondegenerate 
$D$-dimensional lattice:
\beqarn	\label {torusdef2}
	T^D &=& {{\Bbb R}^D \over 2\pi \Lambda}~, \nono
	\Lambda &\equiv& \{ e^a_i m^i; m^i \in \intZ \}~.
\eeqarn
The momentum $k^a$ conjugate to the coordinates $X^a$ on the torus  
 is quantized so that
\[	k \cdot \Delta X \in 2\pi \intZ,	\]
where $\Delta X \in 2\pi \Lambda$ is a lattice vector\myftnote
	{More precisely, eigenstates of $k$ have wavefunction $e^{ik\cdot x}$ 
	in the basis 
	in which $X$ is diagonal.  The wavefunction of a scalar particle 
	must be single-valued.  Since $X$ and $X+\Delta X$ represent 
	the same point of the torus, $e^{\imath k \cdot X}$ must be equal to 
	$e^{\imath k \cdot (X+\Delta X)}$.}.  Therefore 
\[	k^i =  G^{ij} n_j, \come n_j \in \intZ. \]
Namely the momentum $k$ is in the dual lattice $\Lambda^*$ of $\Lambda$,
\[	k^a \in \Lambda^*,	\]
\[	\Lambda^* \equiv \{ e^{*ai} n_i; \come n_i \in \intZ \}, \sepe
		e^{*ai} \equiv e^a_j G^{i j}. \]

Let us consider string compactification over $K=T^D$, with vanishing 
$B$ for the time being, 
\beq	\label {simpleact}
	S_K = \frac 1 {4 \pi} \myI {d^2z} \delta_{ab} \pa X^a \bar \pa X^b.
\eeq
The most general solution of the equation of motion for $X$ is 
\beqar	\label {xmodetorus}
	X^a &=& x^a + 2 p^a \sigma^0 + \omega^a \sigma^1 
			+ \sum_{n \neq 0} \frac \imath n 
				\{ \alpha_n^a e^{-\imath n \sigma^+}
			+ \tilde \alpha_n^a e^{+\imath n
\sigma^-} \} \nono
	&=& x^a - \imath p^a (z + \zbar) - \imath \omega^a (z - \zbar) / 2
		+ \sum_{n \neq 0} \frac \imath n 
			\{ \alpha_n^a e^{- n z}
				+ \tilde \alpha_n^a e^{-  n \zbar}
				\}.
\eeqar
This differs from the solution for $\realR^d$ (\ref {xmode}) 
in a new term linear in $\sigma^1$.  As $\sigma^1$ goes from 
$0$ to $2\pi$, $X^a$ is displaced by $2\pi \omega^a$.  On a 
$T^D$, $\omega^a$ does not have to vanish, 
because the closed string can wind around 
a nontrivial loop on $T^D$, provided $\omega^a \in \Lambda$.  
For this reason $\omega^a$ is called 
 the \emph{winding number}.
 After canonical quantization, we find the same commutation relations 
between $x^a$, $p^a$, and the mode operators as in the last lecture.  
The $\omega^a$'s commute among themselves as well as with the other operators.  
So we can group states into 
\emph {winding sectors} --- 
eigensubspaces of $\omega^a$.
On the other hand, since $p^a$ is the momentum conjugate to the 
center of mass position $X^a$, by our previous discussion  
it takes values in $\Lambda^*$.

	Let us rewrite the mode expansion 
for $X$'s in a more symmetrical form:
\[
	X^a = X^a_L + X^a_R;
\]
\beqar
	X^a_L = x^a_L  -\imath p^a_L z 
			+ \sum_{n \neq 0} \frac \imath n \alpha_n^i e^{-n z},	
			&\sepe&	X^a_R = x^a_R - \imath p^a_R \zbar 
			+ \sum_{n \neq 0} 
			\frac \imath n \tilde \alpha_n^i e^{- n \zbar};	\nono
	x^a_L = \frac 1 2 x^a - \theta^a; 
	&\sepe& x^a_R = \frac 1 2 x^a + \theta^a; 
	\label {plr} \\
p^a_L = p^a + \frac {\omega^a} 2 = e^{* a i} n_i + \frac 1 2 e^a_i m^i; 
&\sepe& p^a_R = p^a - \frac {\omega^a} 2 = e^{* a i} n_i - \frac 1 2 e^a_i m^i. 
\nonumber
\eeqar	
Here we have introduced operators $\theta^a$ which are the canonical conjugates
to the winding numbers $\omega^a$.  Their existence is ensured by 
the existence and uniqueness of a winding sector for each winding number.
$p_L$ and $p_R$ are called \emph {left} and \emph {right momentum} respectively.  
They also appear in the energy-momentum tensor:
\[	L_n = \sum_m \frac 1 2 \nod {\alpha_{n-m} \alpha_m}, \come
\alpha_0^a = p_L^a	\]
\[	\tilde L_n = 
	\sum_m \frac 1 2 \nod {\tilde \alpha_{n-m} \tilde \alpha_m}, \come 
			\tilde \alpha_0^a = p_R^a	\]
The OPE between the vertex operators $e^{i k\cdot X_L}$ is
\beq 	\label {leftope}
	\nod {e^{\imath k_1 \cdot X_L(z)}} \nod {e^{\imath k_2 \cdot X_L(w)}}
	\sim (z-w)^{k_1 \cdot k_2} 
		\nod {e^{\imath k_1 \cdot X_L(z) + \imath k_2\cdot X_L(w)}}
   ~,
\eeq
with similar expression for the right movers.

	The expressions for $p_L$ and $p_R$ suggest some ``duality'' between the 
winding number $\omega$ and the momentum $p$.  
Consider a pair of 
compactification lattices whose lattice vectors $e^a_i$ and
$e'^a_i$ are related as $e'^a_i = 2 e^{*ai}$.  These two
compactifications give the same spectrum since
their allowed values of the momenta are related as
\beq	\label {tdualp}
		p_L \bij p_L';\sepe p_R \bij -p_R'
\eeq
by interchanging the labels $n_i$ and $m_i$. 

\subsection {Example: Compactification on $S^1$}

	Let us try out the above construction
on the simplest case: compactification 
over a circle of radius $R$.  Then the lattice structure is trivial:
\[	e^1_1 = R; \sepe e^{*1}_1 = \frac 1 R.	\]
The allowed values for the momenta are simply
\beq	\label {plrs1}
	p_L= \frac 1 R n + \frac R 2 m, \sepe 
	p_R = \frac 1 R n - \frac R 2 m.
\eeq
The duality just mentioned also takes a simple form.  Consider 
another theory compactified on radius $R' = \frac 2 R$.  If we 
interchange $n$ and $m$ in (\ref {plrs1}), then 
we can identify the momentum operator for $R' = \frac 2 R$ with 
that of $R$ with the isomorphism (\ref {tdualp}).  Now 
extending this to an isomorphism of the fields in the two theories, 
the commutation 
relation between $x_{L,R}$ and $p_{L,R}$ forces us to require 
also 
\beq	\label {tdualx}
		x_L \bij x_L';\sepe x_R \bij - x_R'.
\eeq	
In order to have the spacetime interpretation of this duality 
as inverting the radius of (or equivalently the metric $G_{ij}$ on) 
the circle, we need to transform the oscillators as well:
\beq	\label {tdualosc}
		\alpha_n \bij \alpha_n'; \sepe
		\tilde \alpha_n \bij \mathbf {-} \tilde \alpha_n'.
\eeq
The isomorphism (\ref {tdualp}, \ref {tdualx}, \ref {tdualosc}) can be 
summarized in a more compact form\myftnote
	{As a side remark, we note that this is a two-dimensional version of the 
	``electro-magnetic'' duality discussed in the later courses of
        this school.  }:
\beq	\label {tdualX}
	X_L \bij X_L';\sepe X_R \bij - X_R'.
\eeq
This isomorphism of operators clearly translates into an isomorphism of the 
Hilbert space.  To see this more explicitly, we can compute the partition 
function (\ref {parti}):
\beqar	
	Z &=& \Tr q^{L_0 -  1 / 24} \bar q^{L_0 - 1 / 24} ,	\nono 
	&=& \frac 1 {\left| q^{ 1 / 24} \prod^\infty_{n = 1} (1-q^n) \right| ^ 2}
		\sum_{n, m} q^{\frac 1 2 (n / R + m R / 2)^2}
					\bar q ^ {\frac 1 2 (n / R - m R / 2)^2}, \nono
		&& q = e^{2\pi \imath \tau},\come \im \tau \geq 0.
	\label {s1parti}
\eeqar
	
It is invariant under $R \to \frac 2 R$ \myftnote 
	{One can also evaluate the path integral on closed Riemann surfaces 
	of arbitrary genus.  There $R \to \frac 2 R$ is an invariance provided 
	one shifts the constant dilaton field appropriately.  See \cite {tdual}.}. 
To show that the two theories are actually equivalent, we 
have also to show that this map is an operator algebra isomorphism.  This 
is easy, since both theories are free and their operator product expansions 
can be computed exactly.  Thus $R \to \frac 2 R$ is an exact symmetry 
of the action (\ref {cfactorus}), on arbitrary Riemann surfaces.  
But is it really a symmetry of the spacetime theory 
that the worldsheet action describes?  From the 
discussion of string perturbation in lecture one, 
we see that it is  
a symmetry of string theory order by order in string perturbation expansion.  
In fact, as we will see presently,  
it is a \emph {gauge symmetry} of string theory.

\subsection {Self-Dual Radius: $R = \protect \sqrt 2$}

At the particular value of $R = \sqrt 2$\myftnote 
	{If we put $\alpha'$ back, the self-dual radius would be 
	$\sqrt {\alpha'}$ by dimensional analysis.}, the duality
$R \to \frac 2 R$ maps $R$ back to its original value and we expect 
something interesting to occur.  Indeed, at this radius, the partition 
function (\ref {s1parti}) can be rewritten, after some elementary 
manipulation, as 
\beq	\label {sfduparti}
	 Z = \left| \frac 1 {\eta} \sum_n q^{n^2} \right| ^2
		+ \left| \frac 1 {\eta} \sum_n q^{(n + 1 / 2)^2} \right|^2,
\eeq
where Dedekind's $\eta$-function is simply the denominator in (\ref {s1parti}):
\[	\eta \equiv 
	q^{\frac 1 {24}} \prod^\infty_{n = 1} (1-q^n).	\]
The first term in the sum is the modulus squared of 
$q^{- \frac 1 {24}} (1 + 3q + \cdots)$.
The second term is that of $q^{- \frac 1 {24}} (2 q^{\frac 1 4} + \cdots)$.  
They (the unsquared sums) are actually the character formulae 
of the two irreducible 
representation of an algebra, called $SU(2)$ affine Lie algebra at level 
$k = 1$.  Where does the algebra comes from?  

	The contributions from (\ref {sfduparti})
to the massless spectrum of the complete string theory are those terms 
first order in $q$ and $\bar q$ inside the absolute values signs.  
Let us look at the left movers.  The states are 
\[	\alpha_{-1} \ket {p_L = 0}, ~~| {p_L = \pm \sqrt 2} \rangle,
       	\]
which respectively correspond to vertex operators
\beq	\label {internalm0}
	\pa X_L, ~~ e^{\pm \imath \sqrt 2 X_L}.
\eeq
The first of the three exists for arbitrary radius $R$, but it is not 
difficult to show that the last two states only exist in the spectrum 
when $R= \sqrt 2$.  
Using (\ref {usefulope}) and (\ref {leftope}),
one can evaluate the OPE's among them as
\beq	\label {kac-moody}
	J^a(w) J^b (z) \sim \frac {k \delta^{ab} / 2} {(w-z)^2}  
		+ \frac {i {\epsilon^{ab}}_c J^c(z)} {(w-z)},	
\eeq
\[	
	J^1 \equiv \half (e^{\imath \sqrt 2 X_L} + e^{- \imath \sqrt 2 X_L}),\come 
	J^2 \equiv \frac {- \imath} 2
		(e^{\imath \sqrt 2 X_L} - e^{- \imath \sqrt 2 X_L}), \come
	j^3 \equiv \imath \frac 1 {\sqrt 2} \pa X,
\]
with $k = 1$.  Here $\epsilon^{ab}_c$ is the structure constant 
of $SU(2)$.  This is precisely the definition of $SU(2)$ affine Lie 
algebra with level $k=1$.  The same story is repeated for the right movers.  

To construct a consistent bosonic string theory, we have also to take 
into account $M^{25}$, the external part of the spacetime.  
Let the coordinate for the $S^1$ be $X^{25}$.  To 
make a vertex operator for the massless particle 
we have to choose, for both the left 
and right moving parts, contributions from either the coordinates on 
$M^{25}$  or $X^{25}$.  At generic radius, they are 
\[	\pa X^\mu \bar \pa X^\nu, \come \mu, \nu = 0, \ldots, 24,		\]
which include the graviton, the antisymmetric tensor, and the dilaton 
in $M^{25}$,  
\[	\pa X^\mu \pb X^{25}, \sepe \pa X^{25} \pb X^\mu,	\]
which correspond to two $U(1)$ gauge fields\myftnote 
	{Weinberg showed in \cite {weinberg} that a consistent theory 
	of massless spin one field $A_\mu$ must have gauge invariance with 
	$A_\mu$ as gauge field.}
 in $M^{25}$, and 
\[	\pa X^{25} \pb X^{25}, \]
a neutral scalar field in $M^{25}$.  This is the usual massless part of 
Kaluza-Klein spectrum.  But as we just found, at $R=\sqrt 2$, 
there will be two additional gauge fields, 
\[	\pa X^\mu e^{\pm \imath \sqrt 2 X^{25}_R}, \sepe
	e^{\pm \imath \sqrt 2 X^{25}_L} \bar \pa X^\mu	\]
and $8$ more scalars 
\[	
	e^{\pm \imath \sqrt 2 X^{25}_L} e^{\pm \imath \sqrt 2 X^{25}_R}, \sepe 
	\pa X^{25}_L e^{\pm \imath \sqrt 2 X^{25}_R}, \sepe 
	e^{\pm \imath \sqrt 2 X^{25}_L} \pa X^{25}_R.
\]
Thus we have in total $6 = 3 + 3$ gauge 
fields and $1 + 8 = 9 = 3 \times 3$ scalars at the massless level.  
The OPE (\ref {kac-moody}) 
allows one to calculate tree level 
S-matrix elements among the gauge fields, and one finds that 
the gauge group is $SU(2)$.  
Hence we see that, at $R = \sqrt{2}$, we have 
an enhancement of gauge symmetry from $U(1)_L \times U(1)_R$\myftnote 
	{Here the subscript ``L'' (``R'') just refers to their origin 
	from the left (right) movers. It has nothing to do with 
	spacetime chirality.} 
to $SU(2)_L \times SU(2)_R$ with a Higgs transforming in $(3,3)$ under 
them.  
To make 
a connection between this observation and the $R \to \frac 2 R$ 
duality, let us make a digression into conformal field theory.

	From the last lecture we see that conformal invariance and hence cancellation 
of conformal anomaly is crucial for 
a consistent string theory.  Generic conformal field theories do 
not have a spacetime interpretation.  Since only the spacetime in 
the uncompactified Minkowski space is observable, one may consider  
using arbitrary CFT to represent the effects of 
``compactification'' even if they do not have any spacetime interpretation 
like that of (\ref {cfactorus}).  This is consistent 
as long as they have the right amount 
of central charge so that the total conformal 
anomaly still cancels.  Therefore 
we should study the moduli spaces of CFT.  
Recall that for a given conformal field theory, we may perturb it by adding 
\emph {marginal} operators
to the action while maintaining conformal invariance.  Therefore 
the space of marginal operators for a theory at a particular 
point on the moduli space of CFT is the tangent space at that point.

A marginal operator for two-dimensional conformal field theory is 
one with conformal dimension $(1, 1)$\myftnote
	{In general, this is just a necessary condition.  One must also require 
	that after an infinitesimal 
	deformation by themselves their conformal weights remain unchanged.}.  It 
is not too difficult to see that  
they are exactly those which create scalar massless particles in the 
Minkowski space.  Therefore, for our case, there seem to be $9$ 
independent directions 
to deform the $c=1$ conformal field theory away from the self-dual 
radius, corresponding to giving VEV's to the $(3,3)$ Higgs.  When the 
VEV is turned on,  
the $SU(2)_L \times SU(2)_R$ 
gauge symmetry is spontaneously 
broken down to a $U(1)_L \times U(1)_R$.
However, the same gauge symmetry tells us we can always choose a gauge 
so that only one component of them, say, the one coupled to 
$\pa X^{25} \pb X^{25}$, has 
a nonzero VEV $a$.  It has the simple spacetime interpretation of 
$a = R-\sqrt 2$ when $R$ is close to the self-dual value $\sqrt{2}$.  
Moreover, there is 
a residual $Z_2$ gauge symmetry, namely the Weyl group for either 
of the two $SU(2)$'s, which inverts the sign of the $a$.  
This is to first order the map 
	\(R = (\sqrt 2 + a) \to \frac 2 R = \sqrt 2 - a + O(a^2) \).
Therefore at generic radius $R$, the T-duality $R \to \frac 2 R$ 
is the remnant of a spacetime gauge symmetry.

	From this we also see that near the self-dual radius, 
the moduli space for the conformal field theory 
corresponding to compactification on a circle is one-dimensional 
and looks like figure 8.
\myftnote
	{The complete moduli space is much more complicated, with a new branch 
	and some discrete points.  See \chap 8.7 of \cite {ginsparg} 
	for an introduction.}
If we try to go to smaller radius than $\sqrt 2$, we will end up, via 
T-duality, with a larger radius.  This is a hint that 
string theory possess a minimal length scale $\sqrt {\alpha'}$
and we cannot probe or define the physics at a smaller scale\myftnote 
	{This statement requires significant qualification after D-branes come 
	into the story, as discussed by S. Shenker in this school.}.

\myepsf {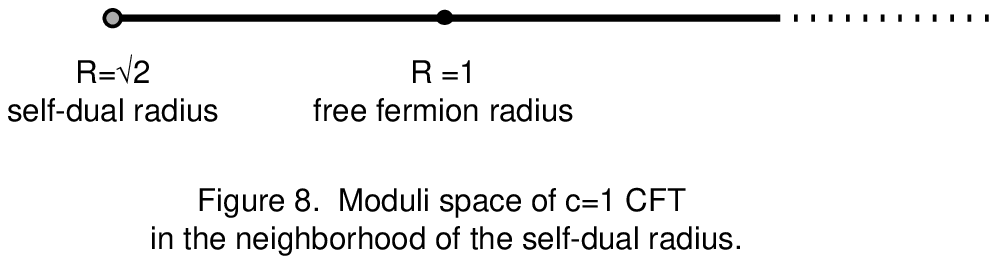}
		
\subsection {$R = 1$}

	Let us also study the case $R = 1$ or 
equivalently $R = 2$.  The motivation will be self-evident 
soon.  By using the product formulae of the \emph {theta functions},
we can rewrite the sums in the partition function into
products:
\beqar	\label {r1parti}
	Z 
	&=&\half \brac { \abs {\frac 1 \eta \sum_n q^{1 / 2 n^2} }^2 
				+ \abs { \frac 1 \eta 
						\sum_n (q^{\frac 1 2 n^2} (-1)^n) }^2
				+ \abs {\frac 1 \eta 
						\sum_n q^{\frac 1 2 (n+\frac 1 2)^2} }^2 }	\nono 
	&=& \half \abs {q^{- \frac 1 {24}}}^2 
\left\{ \abs {\prod_{r=1}^\infty (1 + q^{r - \frac 1 2})^2 }^2 
		 + \abs { \prod_{r=1}^\infty (1 - q^{r - \frac 1 2})^2 }^2
				\right.	\nono
		&& \left.  + \abs { 2 q^{\frac 1 8} 
		 			\prod_{r=1}^\infty (1 + q^r)^2 }^2 \right\}.
\eeqar
We will recover the same partition from worldsheet 
\emph {fermions}.

The spectrum of momenta at $R=1$ is 
\beq	\label {s1lattice}
	p_L = n + \frac 1 2 m; \come p_R = n - \frac 1 2 m, \come m, n \in \intZ.
\eeq
To understand why this is related to the fermions, consider
the following operators
\beqar	\label {fermionop}
	\Psi_L (z) = e^{\imath X_L (z)},&\sepe& 
	\bar \Psi_L (z) = e^{- \imath X_L (z)} \nono 
	\Psi_R (\zb) = e^{\imath X_R (\zb)},
		&\sepe& \bar \Psi_R (\zb) = e^{- \imath X_R (\zb)}.
\eeqar
One can calculate the OPE between them:
\beq	\label {freefermop}
	\Psi_L(z) \Psibar_L (w) \sim \frac 1 {(z-w)}, \sepe
	\Psi_R(\zb) \Psibar_R (\bar w) \sim \frac 1 {(\zb-\bar w)}
\eeq
As we will see, these are precisely the OPE's for a massless 
free Dirac fermion $\Psi$
on the worldsheet, with $\Psi_L$ and $\Psi_R$ being its two Weyl 
components.  
Thus we call them free fermion operators. 
To be precise, $\Psi_L$ and $\bar{\Psi}_L$ do not have corresponding
states in the spectrum since they carry momenta $p_L = \pm {1 \over 2}$
and $p_R = 0$ and hence map states labeled by $(n,m) \in \intZ \otimes
\intZ$ to states that are not. What we do have in the spectrum are
operators bilinear in
these fermions as $\Psi_L \Psi_R$, $\Psi_L \bar{\Psi}_R$.  

As
\(\sigma^1 \to \sigma^1 + 2\pi\), there are two types of
boundary conditions for the fermions. 
When $p_L \in \intZ + \half$ (and thus $p_R \in \intZ + \half$),
they obey the periodic, \emph {Ramond} (R),  boundary condition.   
On the other hand, when $p_L \in \intZ$ (and accordingly
$p_R \in \intZ$), the fermions obey the anti-periodic,
\emph {Neveu-Schwarz} (NS), boundary condition\myftnote
	{This statement requires some further elaboration. If 
	\(\nod {\exp{(\imath X_L)}}\) were given by 
	\(\exp{(\sum_{n > 0} (\imath / n) \alpha_{-n}^i e^{ n z})} \)
	\( \exp{(\imath x_L)} \)  \(\exp{(- \imath p_L z)}\)
	\(\exp{( \sum_{n > 0}  (\imath / n) \alpha_{+n}^i e^{- n z})}\),  
	by our definition NS (R) sector 
	should be (anti-)periodic.  
However, in the the coordinate system we are using, where the worldsheet is 
a cylinder, there should be an additional factor of $e^{- z/2}$.  This modifies 
	the term linear in $z$ on the exponent to \(e^{(p_L - 1/2)z}\)
	and gives the correct periodicity.  The simplest way 
	to understand the origin of this factor is to note that 
	the operator $\nod {e^{\imath k X_L(z)}}$, when acting on the 
	vacuum, should create a state with energy 
	$k^2/2$ since this vertex operator's conformal 
	weight is $(k^2/2, 0)$.  In the cylindrical 
	coordinate we are using, it should have a (Euclidean) time dependence of 
	$e^{-k^2 t / 2}$.  Since it is holomorphically dependent on $z$, the 
	correct factor is $e^{-k^2z/2}$. In our case, $k = \pm 1$, and
   the factor is $e^{-z/2}$.}.  
Clearly both types of boundary conditions 
show up in the lattice $(n,m) \in \intZ \otimes \intZ$.   
The free fermion operators do not mix between R and NS 
boundary conditions.  This suggests us to define 
two corresponding sectors of Hilbert space to host the  
free fermion operators.  
The periodic boundary condition happens when
$(n, m) \in \intZ \otimes (2\intZ+1)$. We call them in
the \emph {R-R sector} since both $\Psi_L$ and
$\Psi_R$ obey the R boundary condition.
On the other hand, the anti-periodic boundary 
condition is realized in the \emph {NS-NS sector} 
with $(n,m) \in \intZ \otimes (2\intZ)$. 

Hence the
worldsheet boson describing compactification over a circle of radius $R=1$ 
is \emph {equivalent} to the
worldsheet fermions after including both  
periodic and anti-periodic boundary conditions 
and then take a certain projection (explained below).  
This is called \emph {bosonization} 
or \emph {fermionization} depending on how you look at it.

	Now let us study this from the fermion side.  Consider the action for a
massless Dirac spinor $\Psi(z)$ in $(1+1)$ dimension\myftnote
{In $(1,1)$-dimensional or in $(T, T+8k)$-dimensional spacetime,
the Weyl condition 
can be compatible with the Majorana condition.  For instance, on the worldsheet, 
	which has signature $(1,1)$, one can define 
\(\gamma^0 = \left( \bear {c c}	0 & \imath \\ \imath & 0 \ear \right), \come 
	\gamma^1 = \left( \bear {c c}	0 & - \imath \\ \imath & 0 \ear \right)\) 
	which are purely imaginary.  Then $\imath \gamma^a \pa_a$ is real and it 
	is consistent with Dirac equation to require 
	$\psi$ to be real, i.e. Majorana.  At the same time, 
$\gamma^0 \gamma^1 = \left( \bear {cc} -1 & 0 \\ 0 & 1 \ear \right)$ is diagonal 
	and real as well, so we can define 
	\( \psi = \left( \bear {c} \psi_L \\ \psi_R \ear \right) \), 
	with $\psi_L$ and $\psi_R$ each being a \emph {Majorana-Weyl} fermion.  
	However, such a thing does not exist if the signature is (2,0), so
	to Euclideanize the worldsheet, we should combine pairs of Majorana-Weyl 
	spinors $\psi$'s into complex Weyl spinors $\Psi$'s.}:
\beqar	
	S &=& \frac \imath {2 \pi} \myI {d^2 \sigma} \bar\Psi \gamma^a \pa_a \Psi, \nono
	  &=& \frac \imath {\pi} \myI {d^2 \sigma} \Psibar_L \pb \Psi_L 
		- \frac \imath \pi \myI {d^2 \sigma} \Psibar_R \pa \Psi_R, \nono
		&& \bar \Psi \equiv \Psi^+ \gamma^0.
\eeqar
Just like the bosonic theories we have discussed so far, the left and right 
movers decouple.  In fact, it is a conformal field theory 
with the same central charge $c = 1$. 
Let us concentrate on the left movers.  We will
freely drop 
the subscript ${}_L$ without warning when there is no ambiguity.  
The mode expansions for the fermion fields are
\beq	\label {fermode}
	\Psi(z) = \sum_r \Psi_{r} e^{- r z}; \sepe
	\Psibar (z) = \sum_r \Psi_{r} e^{- r \zbar}	
\eeq
Canonical quantization gives commutation relation between the modes:
\beq	\label {fermcom}
	\{ \Psi_r, \Psibar_s \} = \delta_{r+s, 0},
\eeq
from which one can derive the OPE (\ref {freefermop}) we previously 
calculated through bosonization.  To make contact with 
the two sectors of Hilbert space discussed earlier, we note that the 
fermions are worldsheet spinors.  As such, they can be either 
periodic or anti-periodic as $\sigma^1 \to \sigma^1+ 2\pi$,  
identified as R and NS sector respectively.  
The (anti-)periodicity also determines the modding $r$ in (\ref {fermode}).
Therefore, $r \in \intZ$ in R sector and 
$r \in \intZ + \half$ in NS sector.

Using bosonization, we can also find 
\beq	\label {jbosonized}
	\nod {\Psi_L(z) \Psibar_L(z)} = \imath \pa X_L (z),	
\eeq
where : : denotes the non-singular part 
of the OPE $\Psi_L(w) \Psibar_L(z)$ 
in the limit $w \to z$.  $\pa X_L$ is the current associated with 
the $U(1)$ symmetry which shifts $X_L$ by a constant.  The charge for this 
current is its zero mode $p_L$.  Since the fermion operators $\Psi_L$ and 
$\Psibar_L$ carry $p_L = \pm 1$, the operator 
\[	F_L \equiv p_L	\]
measures the fermion 
number.  The bosonization rule 
(\ref {jbosonized}) allows us to reexpress the energy-momentum tensor 
for the bosonic theory 
in terms of the fermionic fields
\[	T(z) = \sum_n L_n e^{- n z} = - \half :\pa X \pa X: 
		 = - \half :\Psibar \pa \Psi: - \half \nod { \Psi \pa \Psibar} ,
\]
in agreement with the energy-momentum 
tensor for the fermionic theory found by the usual means.
In particular, its zero mode is 
\beq	\label {ferml0}
	L_0 = \sum_{r > 0} \, r\,  (\Psi_{-r} \Psibar_r + \Psibar_{-r} \Psi_r)	
    ~.
\eeq

Now we are ready to compute the partition functions for the fermionic 
theory and have our final check of bosonization.  We have 
for NS sector,
\[	\Tr q^{L_0} = \prod_{r=1}^\infty (1+q^{r-\half})^2,	\]
\[	\Tr (-1)^F q^{L_0} = \prod_{r=1}^\infty (1-q^{r-\half})^2,	\]
and for R sector,
\[	\Tr q^{L_0} = (1 + 1) q^{\frac 1 8} \prod_{r=1}^\infty (1+q^r)^2,	\]
\[	\Tr (-1)^F q^{L_0} = (1 - 1) q^{\frac 1 8} \prod_{r=1}^\infty
(1-q^r)^2	= 0.\]
Some explanation is warranted.  The squares in all four expressions are due to  
that we have both $\Psi_L$ and $\Psibar_L$ at each modding.  
In R sector, since the modding 
is even, we have the commutation relation 
\[	\{ \Psi_{0}, \Psibar_{0} \} = 1.	\]
This is represented by a single fermionic oscillator.  Note we can 
also rewrite this as the Clifford algebra in two dimensions:
\[	\{ \psi_{i}, \psibar_{j} \} = 2 \delta^{ij}, \sepe i, j = 1, 2,	\]
\[	\Psi \equiv \half (\psi^1 + \imath \psi^2).	\]
Its straightforward generalization to higher 
even-dimensional Clifford algebra will be useful in the next lecture.  
The R ground states here therefore consist of two states: $\ket +$ and $\ket -$:
\[	\Psi_{0} \ket {-}=0, ~~\Psibar_{0} \ket {+} = 0,	\]
\[	\Psi_{0} \ket {+} = \ket {-}, \sepe \Psibar_{0} \ket {-} = \ket {+}.	\]
States built from $\ket {+}$ and $\ket {-}$ with nonzero modes 
give identical contribution to $\Tr q^{L_0}$.  
However they have opposite $(-1)^F$ parity as 
\beq	
	\acomm {\Psi_{0}} {(-1)^F } = 0,~~ 
\acomm {\Psibar_{0}} {(-1)^F} =0.
\eeq
Therefore their contributions to $\Tr (-1)^F q^{L_0}$ are 
equal in magnitude but opposite in sign.
The R ground states have charges $p_L = \pm \half$, and their 
conformal weight is $\frac 1 8$ which accounts for the $q^{1/8}$ factor in 
the R sector partition function.  In fact one 
can map the NS ground state to the two R ground states using operators 
\[\sigma^{\pm} \equiv e^{\pm \imath \half \phi_L}	\]
with conformal weight $\frac 1 8$ \myftnote
	{As mentioned earlier, the boundary condition of the left and right 
	moving sector are correlated.  The complete operator that does this is 
	$\exp (\pm \imath \half \phi_L \pm \imath \half \phi_R)$.}.
Using the OPE (\ref {leftope}) one can show that they 
transform into each other under, and flip 
the boundary condition of, the free fermion operators.  
They are called spin field operators and form a representation of 
Clifford algebra in two dimensions under the actions of $\Psi$ 
in OPE.

Combining the left and right movers, we can rewrite (\ref {r1parti}) as
\beq	\label {sumspinparti}
	Z = \Tr_{\mrm{NS-NS \oplus R-R}} \:P\: q^{L_0 - 1/24} \bar q^{\tilde L_0 - 1/24}
\eeq
where we introduce a projection operator 
\beq
	P \equiv \paren {1 + (-1)^{F_L + F_R}} / 2.
\eeq
This $P$ projects out states with odd number of fermions
from NS-NS and R-R sectors to obtain the Hilbert space 
for the original bosonic theory.
 
\subsection {Modular Invariance and Narain's Condition}

Let us continue the discussion of string on $S^1$ with
$R = 1$. 
 As mentioned 
before, the partition function 
\( {\rm Tr} q^{L_0 - 1/24} \bar q^{\tilde L_0 - 1/24}\) for a 2d theory 
can be interpreted as evaluating its path integral 
on a torus (fig. 6).  The
path integral formally involves integrating over all possible 
field fluctuations, which must  satisfy 
appropriate boundary conditions when 
a nontrivial manifold is involved.  The theory
is that of free fermions (spinors).  Spinors cannot be 
defined on all manifolds.  But when they can, there is often more 
than one consistent but inequivalent way to do so, 
called \emph {spin structures}.  
A proper explanation of these matters is 
outside the scope of these lectures but can be found in \chap 12 of 
\cite {GSW}.  As a matter of fact, there are four consistent ways 
to define spinors on a torus, corresponding to choosing periodic or 
anti-periodic boundary conditions as a spinor goes around
either of the two independent 
nontrivial cycles shown in figure 5.  After identifying 
$\gamma_2$ as ``time'' and $\gamma_1$ as the spatial extent of the string, 
R (NS) sector corresponds to (anti-)periodic boundary condition around 
the latter.  Because $(-1)^F$ anticommutes with $\Psi$, its insertion 
in the trace flips the boundary condition around $\gamma_2$.  
It can be shown that without its insertion, that the boundary condition is 
antiperiodic.  These situations are summarized in figure 9. 
Therefore we can interpret the projection and the sum over 
R-R and NS-NS sectors in (\ref {sumspinparti}) as a sum over all possible 
spin structures, 
but what is the reason behind this summation?

\myepsf {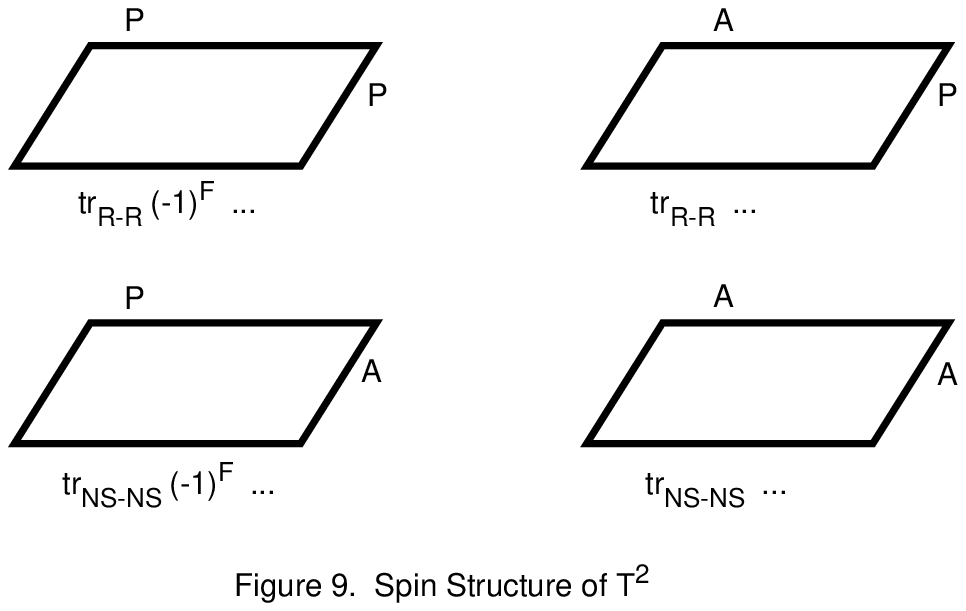}

To understand this we need to introduce the notion of modular invariance.  
As mentioned in the last lecture, we can characterize the shape of a 
torus by a complex parameter $\tau$ taking value in the upper complex
half-plane.  However, not all distinct values of $\tau$ correspond to distinct
tori.  In fact, define the operations $T: \tau \to \tau +1$ and 
$S: \tau \to - \frac 1 \tau$.  
The operation $T$ corresponds to changing 
one of the lattice basis defining the torus
and $S$ to swapping the basis.
They generate \emph {large diffeomorphisms} of the torus, 
which cannot be smoothly connected to the identity map.  
 It should be clear that 
they map the different spin structures into each other.  
Only the spin structure which is periodic around 
both cycles is invariant.  This is not surprising since the 
corresponding partition function vanishes identically.  
For the bosonic theory, 
the counterpart of spin structures correspond to the different 
windings around the target space $S^1$ as the coordinate $X$ 
goes around the two cycles of the worldsheet torus.  When we sum 
over all possible value for the 
center of mass momentum and the winding numbers, we 
are summing over all these different contributions, rendering  
the partition function invariant under $S$ and $T$ --- it 
is \emph {modular invariant}. Therefore to 
have an equivalence between the bosonic and fermionic theories, 
we \emph {must} sum over spin structures on the fermionic 
side.

	As a side note, $S$ and $T$ generates the group \slZ, 
the group of $2\times 2$ 
matrices with integral elements and unit determinant:
\[	T: \left( \bear {cc} 1 & 1 \\ 0 & 1 \ear \right) \]
\[  S: \left( \bear {cc} 0 & -1 \\ 1 & 0 \ear \right)	\]
This group will appear time after time throughout this school\myftnote 
	{It eventually made its way to the official T-shirt for this
school.}.  
Here we merely note that 
they have the interpretation of changing the basis $(e^1, e^2)$ 
of the lattice defining torus:
\[	\left( \bear {cc} a & b \\ c & d \ear \right) \in \slZ:
	\column {{e^1}' \\ {e^2}'} 
	= \left( \bear {cc} a & b \\ c & d \ear \right) \column {e^1 \\ e^2}
\]
Their action on the moduli $\tau$ is 
\[ \tau \to \tau' = \frac {a \tau + b} {c \tau + d}. \]
This discrete identification divides the upper complex plane into infinite 
number of \emph {fundamental domains},
each of which is a single 
cover of the true moduli space for the torus.

We will now demonstrate the modular invariance of 
the partition for the most general class of toroidal 
compactification of the bosonic string.  Recall our earlier expression for 
the left and right moving momenta: 
\[	p^a_L = p^a + \frac {\omega^a} 2 = e^{* a i} n_i + \frac 1 2 e^a_i m^i; 
	\sepe p^a_R = p^a - \frac {\omega^a} 2 = e^{* a i} n_i - \frac 1 2 e^a_i m^i. 
\]
Let us combine them into one $(D+D)$-component column vector:
\[	\hat p = \left( \bear {c} p^a_L \\ p^a_R \ear \right).	\]
This construction treats $\Lambda$ and $\Lambda^*$ on equal footing 
as 
\beq
	\hat p = \hat e^{* i} n_i + \hat e_j m^j,
\eeq
where 
\beq	\label {latticevec1}
	\hat e_j = \frac 1 2
					\left( \bear {c} e^a_j \\ - e^{a}_j \ear \right); \sepe
	\hat e^{* i} = \left( \bear {c} e^{*ai} \\ e^{*ai} \ear \right).
\eeq
Hence $\hat p$ takes value in a $(D+D)$-dimensional lattice $\hat \Lambda$ 
spanned by $\{\hat e^{* i}\}$ and $\{\hat e_j\}$.  

We also define a metric of signature ($D$, $D$) on this $2d$-dimensional 
space:
\[	\hat \delta = \left( \bear {cc} \delta_{ab} & 0 \\ 0 & - \delta_{ab} \ear 
					\right)	\]
This metric captures some of the 
 most important properties of the lattice $\hat \Lambda$.
Because of (\ref {latticevec1}), 
\beq	\label {latticeinsec}
	\hat e_i \cdot \hat e_j = 0; \sepe
	\hat e^{* i} \cdot \hat e^{* j} = 0; \sepe
	\hat e_i \cdot \hat e^{* j} = \delta_i^j,
\eeq
\beq	(\hat e^{* i} n_i + \hat e^j m^j) \cdot 
		(\hat e^{* i} n_i' + \hat e^j {m^j}') = n_i m^{i'} + n_j' m^j
\eeq
which implies (1) if $q \in \hat \Lambda$, then
$q \cdot q \in 2 \intZ$ and (2) the dual lattice of
$\hat \Lambda$ is $\hat \Lambda$ itself.
Such a lattice is called even, because $q^2$ is 
\emph {even}, \emph {self-dual}, because 
$\hat \Lambda^* = \hat \Lambda$, and \emph {Lorentzian}, because of the signature 
of the metric with respect to which the conditions are imposed.

Now consider as the internal part of string ``compactification'' 
a conformal field theory the same as that of (\ref {simpleact}) 
except that its 
momenta live on some general $(D+D)$-dimensional 
lattice $\hat \Lambda$.  Its partition function is 
\[	Z_{\hat \Lambda} = \frac 1 {\abs {\eta (q)}^{2D}}
		\sum_{(p, \tilde p) \in \hat \Lambda} q^{\half p^2} 
								\bar q^{\half \tilde p^2}
\]
What requirement should we impose on $\Lambda$?  Recall that string 
theory has worldsheet diffeomorphism invariance.  Modular transformations 
are diffeomorphisms of the torus that cannot be contracted smoothly to the
identity.  They are residual gauge symmetries after 
gauge fixing.  Like the Weyl rescaling symmetry, it might be 
anomalous quantum mechanically. 
In the last lecture, we see that 
conformal invariance at the quantum level 
is responsible for removing unphysical 
states from the string spectrum and determining the critical dimension.  
Similarly, modular invariance would imply we need only to 
integrate over a fundamental domain as the moduli space 
of the torus.  This turns out to be essential for 
preventing ultraviolet divergences 
in string theory (\chap 8.2 of \cite {GSW}). 
It is natural to ask what kind of $\hat{\Lambda}$ would 
ensure the modular invariance of $Z_{\hat{\Lambda}}$.  

	Since the modular group SL$(2,\intZ)$ 
is generated by $S$ and $T$, 
it is sufficient to require that $Z$ is invariant under both of them.
For the $T$-transformation, 
\[
	Z_{\hat \Lambda}(\tau + 1) = \frac 1 {\abs {\eta}^{2D}} \sum_{(p, \tilde p) \in \Lambda}\
		q^{\half p^2} \bar q^{\half \tilde p^2} 
			e^{2 \imath \pi \half (p^2 - \tilde p^2)}.
\]
Since $(p^2 - \tilde p^2) \in 2\intZ$ for an even Lorentizan lattice
$\hat \Lambda$, the partition function is invariant
under $T$. For the $S$-transformation, we
make use of the \emph {Poisson resummation} 
(see the appendix), 
and the identity 
\[	\eta(- \frac 1 \tau) = \sqrt {- \imath \tau} \eta (\tau).	\]
One then finds
\[ Z_{\hat \Lambda}(- \frac 1 \tau)
		= \vol (\hat \Lambda^*)  Z_{\hat \Lambda^*} (\tau). \]
Here  $\vol(\hat \Lambda)$ denotes the volume of 
the unit cell of the lattice $\hat \Lambda$.
Since 
\[	\vol(\hat \Lambda) \vol(\hat \Lambda^*) = 1	\]
for any lattice $\hat \Lambda$,
$\vol (\hat \Lambda^*) = 1$ provided $\hat \Lambda$ is self-dual.\
In this case, the above equation gives
\[  Z_{\hat \Lambda}( - \frac 1 \tau) = Z_{\hat \Lambda} (\tau) . \] 
Therefore if $\hat \Lambda$ is an even self-dual Lorentzian 
lattice, $Z_{\hat \Lambda}$ is modular invariant and is a 
candidate for consistent string compactification.  This is known as 
\emph {Narain's condition} (ref. 340 in \cite{GSW}, Vol 1).

Let us now figure out what is the moduli space of such
a compactification.  
Any nontrivial $O(D,D)$ rotation 
would map one even self-dual Lorentzian 
lattice into a different one.  
The converse is a mathematical fact:  
any two even self-dual Lorentzian 
lattices are related by some $O(D,D)$ 
rotation.  Therefore the space of such lattices is simply $O(D,D)$.  
However, not all of them correspond to different compactifications.  
The spectrum for the $(26-D)$-dimensional theory is determined 
by $p_L^2$ and $p_R^2$.  They are left invariant by  
$O(D)\times O(D)$, the maximal compact subgroup of $O(D,D)$, 
acting independently on the left and right momenta respectively.  
Therefore 
the space of vacua is \emph {locally} $O(D,D) / (O(D)\times O(D))$, 
of dimension $D^2$.  

In fact a \emph {spacetime} interpretation can be given 
to such a construction.  In (\ref {simpleact}) we have set 
to zero the background antisymmetric tensor field $B$.  The 
dimension of the space of possible $G$ is only $D(D+1)/2$.
However $B$'s contribution to the total energy vanishes as long 
as its field strength $H$ is zero.  This allows us 
to give to $B$ arbitrary constant VEV while staying in the  
vacua.  Since $B$ contains 
$D(D-1)$ independent components, this fully accounts for 
the dimension of the space of vacua.  Indeed,  
by canonically quantizing the action 
\[	
	S_K = \frac 1 {4 \pi} \myI {d^2z} 
			(G_{ab} + B_{ab})\pa X^a \bar \pa X^b,
\]
one finds that (\ref {latticevec1}) is modified:
\beq	\label {latticevec2}
	\hat e_j = \half
					\column  {e^{*ai} B_{ji} + e^a_j \\ 
						   e^{*ai} B_{ji} - e^{a}_j }; \sepe
	\hat e^{* i} = \column {e^{*ai} \\ e^{*ai}}.
\eeq
It is easy to verify that this satisfies Narain's condition.

	Just as for the moduli of the worldsheet torus, there are further 
discrete identifications of points in the moduli space of a toroidal
compactification.  Let us now find what they are.
The toroidal compactification does not 
affect the oscillators, and the operator algebra works out as usual.  
All that distinguishes 
one compactification from another is the lattice $\hat \Lambda$ in which 
the left and right momenta live.  
Thus we arrive at the important conclusion that 
any two toroidal compactifications are \emph {equivalent}
if their lattices are the same, i.e. they differ only
by a change of lattice basis.
The most general change of basis is 
an element of SL$(2D, \intZ)$, acting on the labels of lattice basis. 
But when we parameterize the space 
of even self-dual Lorentzian lattices acting by $O(D,D)$ on 
some reference lattice of the form (\ref {latticevec1}), 
the inner product matrix of the basis vectors always takes the 
\( \paren {\bear {cc} 0&I \\ I&0 \ear } \) form in 
(\ref {latticeinsec}).  Therefore the analog of the modular group 
for the vacua is contained in $O(D,D;\intZ)$,  
the stabilizer of (\ref {latticeinsec}) in SL$(2D, \intZ)$.  
It is easy to identify some of its elements.  For example, the 
analog of $T: \tau \rightarrow \tau +1$ 
is adding to $B_{ij}$ an integral 
antisymmetric matrix.  The analog of $S: \tau \rightarrow -1/\tau$ 
is to change the basis of the compactification lattice $\Lambda$.  And then 
there are the generalizations of the $R\to \frac 2 R$ symmetry.  Since 
$T^D \sim (S^1)^D$, there are now $D$ of them.  
Consideration in a similar vein to that for $R\to \frac 2 R$ 
duality shows that they are \emph {gauge} symmetries.
The detailed forms of these 
discrete transformations can be found in \cite {tdual}.  
They do not commute with each other but actually generate 
the whole $O(D,D;\intZ)$, the \emph {T-Duality} group for 
compactification over $T^D$.  The moduli space for 
such compactifications can therefore be written as 
$O(D,D;\intZ) \bs O(D,D) / (O(D) \times O(D))$.

\subsection* {Appendix: Poisson Resummation}

	Consider a function $f$ defined on $\realR^n$ and its Fourier transform 
$f^*$:
\[	f(x) = \myI {\frac {d^n k} {(2\pi)^n}} e^{\imath k \cdot x} f^*(k).\]
Let $\Lambda$ be some lattice in $\realR^n$ and $\Lambda^*$ be its dual, then 
one finds
\beqarn
  \sum_{m\in \Lambda} f(m)
		& = & \myI {\frac {d^n k} {(2\pi)^n}} f^*(k) 
			\sum_{m\in \Lambda} e^{\imath k \cdot m} \\
		& = & \vol(\Lambda^*) \sum_{n\in \Lambda^*} f^*(2\pi m).
\eeqarn

\section {Lecture Three: Superstrings}

	The bosonic string theory we studied in the last two lectures 
has displayed some very interesting structures, yet it conspicuously 
lacks one important ingredient: fermions.  In the real world, we of course 
know that fermions are the basic constituents of matter.  So we should 
find some way to incorporate them into string theory if the latter 
is to become a theory of reality.  From the last lecture, we see that 
under certain conditions a theory of bosons can be equivalent to 
a theory of fermions.  That was in the context of worldsheet, 
while what we really want are spacetime fermions.  
However the two are not unrelated.  As we have seen, the Ramond 
sector of a theory of worldsheet fermions furnishes a representation of 
the Clifford algebra with the worldsheet fermion operators, 
which carry spacetime 
Lorentz indices, playing the role of 
gamma matrices.  In this lecture we will indeed 
see how to build a theory of \emph {spacetime} 
fermions out of \emph {worldsheet} fermions with worldsheet 
supersymmetry.  At the end of the day 
we will find that the annoying tachyon has disappeared.  Moreover
we will find a symmetry between 
\emph {spacetime}  bosons and fermions.

\subsection {From Superparticle to Superstring}

	Let us start at a more humble level and try to construct an 
action for a superparticle by adding new fields to the worldline action 
for the point particle (\ref {ptact}).  In fact there is more than one way 
to do it, but we will consider what is called 
the \emph {spinning particle}\myftnote
	{Another approach, which exhibits spacetime fermion and supersymmetry 
	manifestly, can also be generalized to string theory
         --- the Green-Schwarz 
	action.  We will mention it briefly below.  For more details, see 
	\chap 5 of \cite {GSW}}.
	
First let us write an action with a worldline einbein
\beq	\label {polypt}
	S = - \half \myI {d\sigma} \brac {e \dot X^2 - \frac {m^2} {e}}.
\eeq
This is to (\ref {ptact}) what the Polyakov action is to the Nambu-Goto action.  
The einbein $e$ is a Lagrange multiplier rather than a dynamical variable.  
By solving equations of motion for $e$ and substituting the solution
back to (\ref {polypt}),
we regain (\ref {ptact}) for $m\neq 0$.  For $m=0$ the latter fails 
but (\ref {polypt}) is still valid as an action for
a massless particle.  Let us supersymmetrize the action (\ref {polypt}) 
when  $m=0$\myftnote
	{If the \emph {cosmological constant} $m \neq 0$, 
	worldline supersymmetry, if present at all, must  
	be spontaneously broken.  To keep the action 
	supersymmetric one must introduce an additional 
	fermion as the Nambu-Goldstone particle, which decouples
	from the rest of the theory in the limit $m \to 0$.  
	We will not consider this case, since for string theory, 
	Weyl rescaling invariance forces $m$ to vanish even for 
	the bosonic string.}.
We add worldsheet Majorana
spinors $\psi^\mu$ as the superpartners of $X^\mu$ and a
worldsheet gravitino 
$\nu$ as the superpartner of $e$.
\beq	\label {superpt}
	S = -\half \myI {d\sigma} \brac {e \dot X^2 
		- \imath e \psi \dot \psi 
		- 2 \imath \nu \dot x \psi}.
\eeq
Clearly this action is invariant under worldline diffeomorphisms.  
As implied above, it also has a local supersymmetry:
\[	\delta X^\mu = \imath \theta \psi^\mu; 
		\sepe \delta \psi = \theta \dot X^\mu; \]
\[	\delta e = - 2 \imath \theta \nu; 
		\sepe \delta \nu = \dot \theta e - \half \theta \dot e. \]
Just as in the Polyakov action, $\nu$ and $e$ do not have dynamical degrees of 
freedom.  Their equations of motion are algebraic and serve to 
impose constraints on the physical phase space.  
Variation of the action with respect to $\nu$ implies 
\beq	\label {constrsupt}
	\dot X \cdot \psi = 0.
\eeq
Canonical quantization for this action yields 
\[	- \imath \frac \pa {\pa X^\mu} = P_\mu = g_{\mu \rho} \dot X^\rho,	\]
\[	\acomm {\psi^\mu} {\psi^\rho} = g^{\mu\rho}.	\]
Therefore $\psi^\mu$ realizes the Clifford algebra 
for the spacetime, for 
which the Hilbert space forms a representation.  The spinning 
superparticle is a spacetime spinor.  The constraint equation (\ref {constrsupt}) 
then states that physical states satisfy the Dirac equation as befit 
for a spinor:
\[	\imath \gamma^\mu \pa_\mu \langle {x} | phys
 \rangle = 0.		\]

	It is simple to generalize this to strings.
The supersymmetrization of the bosonic Polyakov action 
(\ref {polyakovact}) is:
\beqarn 
S = \frac 1 {4 \pi \alpha^\prime}
        \myI {d^2\sigma}
        \sqrt {- g}&& \!\!\!\!\!\!\!\!
     \Big\{ g^{a b} (\pa_a X^\mu \pa_b X_\mu
        + \imath \bar \psi_\mu \lambda^a \pa_a \psi^\mu)
         \nono
   && \!\!\!\!\!\!\!\!
  + \bar \chi_a \lambda^b \lambda^a (\pa_b X^\mu
        + \half \bar \psi_\mu \psi^\mu \chi_b ) \Big\}~.
\eeqarn

Here $\lambda^a$ are the worldsheet Dirac matrices.  
New field contents include 
$D$ worldsheet spinors $\psi^\mu$ that transform 
in spacetime as a tangent vector, and 
a worldsheet Rarita-Schwinger field $\chi_a$ with no spacetime index.
The action has 
four local symmetries: the worldsheet diffeomorphism and Weyl rescaling 
symmetries already present for the bosonic string, 
and their superpartners: local supersymmetry and super-Weyl 
transformation.  Classically they together allow one to gauge away 
the metric $g$ and the Rarita-Schwinger field $\chi_a$, and impose 
constraints on the physical phase space.  
In the \emph {superconformal} gauge, $g_{ab}$ can be set to 
$\lambda \gamma_{ab}$ 
and $\chi_a$ to $0$.  Again, there are 
potential anomalies.  The new 
Faddeev-Popov ghosts introduced 
by gauge fixing the local fermionic symmetries 
raise the central charge for the ghost sector to 
$-15$.  On the other hand, the contribution 
from the $\psi$'s increases  
the matter sector central charge to $\frac 3 2 D$.
Therefore the critical dimension 
for them to cancel is now $D=10$.  

		Like the conformal gauge, the 
superconformal gauge is preserved by some 
residual gauge symmetries, which 
are called \emph {superconformal transformations}.  
The superconformal gauge action, 
\[	S = \frac 1 {\pi \alpha'} 
		\myI {d^2 \sigma} \brac {\pa_+ X \cdot \pa_- X 
			+ \imath \psi_L \cdot \pa_- \psi_L
			+ \imath \psi_R \cdot \pa_+ \psi_R},	\]
is the supersymmetric extension of (\ref {cfact}). 
It is a superconformal field theory (SCFT), a conformal 
field theory with additional structures and algebra reflecting its 
superconformal symmetry.  Gauge fixing 
them leads us again to light-cone gauge, where 2 directions 
of the oscillatory 
excitations are taken away from both the $X$'s and the $\Psi$'s.  Manifest 
Lorentz covariance is lost but the constraints are explicitly solved.  
Back in the superconformal gauge, the same result should be obtained 
if we impose constraint conditions on the physical states in the same 
way as we did for the bosonic string.  The constraint conditions correspond to 
the vanishing of the matrix elements of $T$ ($\tilde T$), 
left (right) moving energy-momentum 
tensor, and $G$ ($\tilde G$), left (right) moving super-current, between 
physical states.

	As discussed in the last lecture, there are two sectors of Hilbert 
space for a worldsheet 
fermion, with different boundary conditions.  Spacetime 
Lorentz covariance requires all the left (right) moving 
fermions to be in the same sector, but we let the choice 
for left and right movers be independent.  Hence the 
superstring has $4$ sectors: 
NS-NS, NS-R, R-NS and R-R, in contrast to what we did 
in the last lecture.    As usual, left and right moving operators 
decouple, and we will concentrate on the left movers:  
\[	T = \sum_n L_n e^{-nz} = 
		- \half \pa X \cdot \pa X - \half \psi \cdot \pa \psi,	\]
\[	G = \sum_n G_n e^{-nz} = \imath \psi \cdot \pa X.	\]
Because $\pa X$'s have integer 
modding, the modding of $G$ is the same as that of $\psi$'s: 
$r \in \intZ$ in R sector; $r \in \intZ + \half$ in NS sector. 
The superconformal algebra is 
\[
	[L_n, L_m] = (n-m) L_{n+m} + \frac D {8} (n^3 - n) \delta_{n+m, 0}.
\]
\[
	\acomm {G_r} {G_s} = 2L_{r+s} + \frac D 2 (r^2 - \frac 1 4) \delta_{r+s, 0}
\]
\[	\comm {L_n} {G_r} = (\half n - r) G_{n+r}.	\]
The corresponding OPE's can be found in \chap 12 of \cite {lust}.

	We learned from the last lecture that  
the R sector realizes the 
Clifford algebra, therefore they transform as 
spacetime spinors.  We also see that for every 
complex $\Psi$ or, equivalently, 
every pair of real $\psi$, an R sector ground 
state is $\half \times \left(\half\right)^2= {1 \over 16}$ 
higher in 
$L_0$ eigenvalue, its conformal weight, than the NS 
ground state.  Therefore 
it is natural to shift $L_0$ by $- \frac D {16}$ for the R sector
so that the R ground state also obeys $L_0 = 0$.  
With this definition of $L_0$, we have 
\[	[L_n, L_m] = (n-m) L_{n+m} + \frac D {8} n^3 \delta_{n+m, 0},\]
\[
	\acomm {G_r} {G_s} = 2L_{r+s} + \frac D 2 r^2 \delta_{r+s, 0},
\]
for the R sector.
In particular $G_0^2 = L_0$.  This is the  
rigid supersymmetry algebra in 2 dimensions.  From 
a spacetime point view, $G_0$ is the Dirac operator $\dirac \pa,$ 
$L_0$ the d'Alembertian operator $\Box$.  So $G_0^2 = L_0$ translates 
into the identity $\dirac \pa^2 = \Box$.  The constraints 
$T \sim 0$, $G \sim 0$ in particular contain the Dirac and the Klein-Gordon 
equations.

	The constraints on physical states are 
\[	(G_r - b \delta_{r, 0} ) \ket {phys} = 0, ~~  
	(L_n - a \delta_{n,0}) \ket {phys} = 0, \sepe r, n \geq 0.\]
In the R sector, the relation between $G_0$ and $L_0$ implies 
$a = b^2$.  Equivalence with the light-cone gauge spectrum then 
leads to $a = 0$ and $D = 10$,
 as the students should verify.  
Below we briefly demonstrate the procedure for the NS sector.

\paragraph {Ground State}
\beq
	(L_0 -a ) \ket {k}  = 0, \implies m^2 = -k^2 = -2a
\eeq

\paragraph {First excited level}
\[
	(L_0 - a) e_\mu \psi_{-1/2}^\mu \ket{k} = 0 \imply 
		m^2 = - k^2 = 1 - 2 a {\frac {\alpha^\prime} 4}.
\]
\[	G_{1/2} e_\mu \psi_{-1/2}^\mu \ket{k} = 0 \imply k \cdot e = 0.	\]
\[	G_{-1/2} \ket {k} \sim 0 \imply 
		e^\mu(k) \sim e^\mu(k) + \xi k^\mu.\]
These two conditions remove 2 degrees of freedom, in agreement 
with the light-cone gauge, only if 
this level is massless.  Hence $a = \half$\myftnote
	{This agrees with $a = 0$ for the R sector and the relative 
	normalization of $L_0$ for the two sectors if and only if 
	$D = 10 - 2$ --- another consistency check.}
and the tachyon remains, for the time being.

\paragraph {3rd excited level}\ \\ 

For reasons similar to the case of the second excited level of 
bosonic string, we require 
\[	(G_{-3/2} + \gamma G_{-1/2} L_{-1} \ket {k} \sepe - k^2 = 2\]
to be physical in order to have the same number of states as 
in the light-cone gauge.  This implies $\gamma = 2$ and $D = 10$.  
Thus we obtain again that the critical dimension for superstring 
is $10$.

Define 
\beq	\label {fermionstrength1}
	N = \mathbf {-\half} + \sum_{n>0} (n \alpha_{-n} \cdot \alpha_{n} 
		+ (n-\half) \, \psi_{-n+1/2} \cdot \psi_{n-1/2})
\eeq 
for the NS sector and 
\beq	\label {fermionstrength2}
	N = \sum_{n>0} n \,(\alpha_{-n} \cdot \alpha_{n} 
		+ \psi_{-n} \cdot \psi_{n})
\eeq 
for the R sector, and similarly define $\tilde N$ for the right movers.
Then the level matching condition 
is again $N = \tilde N$. The mass shell condition 
can be written as 
\[	m^2 = 2N.	\]

\subsection {Spacetime Supersymmetry}
	If superstring theory has spacetime supersymmetry, then its 
one-loop vacuum amplitude should vanish due to cancellation between 
bosons and fermions.  We know from the 
first lecture that such an one-loop amplitude correspond to 
the partition function of the worldsheet action.  The partition 
function also tells us the spectrum of the theory, and unbroken 
supersymmetry would imply a perfect matching between bosonic and fermionic 
spectra.  As we are working on a closed string theory, we need to 
glue the left and right movers together to obtain a physical 
state or vertex operator.  As the R sector realizes the Clifford algebra, 
spacetime bosons should come from  the
NS-NS and R-R sectors and fermions from the NS-R and R-NS sectors.
Hence the superstring partition function takes 
the form:
\[	Z = (Z_{\mrm{NS}} \bar Z_{\mrm{NS}} + Z_R \bar Z_R) 
		-(Z_{\mrm{NS}} \bar Z_R + Z_R \bar Z_{\mrm{NS}})
	  = \abs {Z_{\mrm{NS}} - Z_{\mrm R}}^2, \]
where $Z_{\mrm{NS}}$ and $Z_{\mrm R}$ are the partition function 
for the (left moving) NS and R sectors respectively.  
For $Z$ to vanish, $\zNS - \zR$ must be zero. However\myftnote 
	{We are calculating in 
	the light-cone gauge, or equivalently we have taken 
	into account the ghost contribution.  We also neglect to write 
	the trivial factor of $(2 \im \tau)^{-5}$ from integrating over 
	momentum.
	It is an instructive exercise for the students 
	to justify the various factors and 
	powers of $q$ based on  the discussion from 
	the last lecture.}
\beqarn	\zNS &=& \Tr_{\mrm{NS}} \: q^{L_0 - 12/24} 	\nono
		&=& \brak {\frac {q^{-1/24}} 
				{\prod_{n=1}^{\infty} (1-q^n)}}^8
			\brak {q^{-1/24} 
			\prod_{n=1}^{\infty} (1 + q^{n-1/2})^2}^4	\nono
		 &=& q^{-1/2} \brak 
		 	{\frac {\prod_{n=1}^{\infty} (1 + q^{n-1/2})}
		 			{\prod_{n=1}^{\infty} (1-q^n)}}^8
\eeqarn
is definitely not equal to 
\beqarn
	\zR &=& \Tr_{\mrm{R}} \: q^{L_0 - 12/24} \nono
		&=& \brak {\frac {q^{-1/24}} 
				{\prod_{n=1}^{\infty} (1-q^n)}}^8
			\brak {2 q^{-1/24} q^{1/8} 
				\prod_{n=1}^{\infty} (1 + q^{n})^2}^4	\nono
		 &=& 2^4 q^{-1/2} \brak 
		 	{\frac {\prod_{n=1}^{\infty} (1 + q^{n})}
		 			{\prod_{n=1}^{\infty} (1-q^n)}}^8
\eeqarn
Therefore as it is there is no spacetime supersymmetry.  In fact, 
the ground state in the NS sector is a tachyon, whereas in the R it 
is massless.  Anyhow, the tachyon's presence would indicate vacuum 
instability, in direct conflict with supersymmetry.   
It is therefore clear that to have spacetime supersymmetry 
we have to truncate the spectrum consistently so that the tachyon 
disappears.  This reminds us of the projection operator $P$ introduced 
in the last lecture to obtain a bosonic theory from a fermionic one.  
However, since we want to remove the tachyon, the projection operator 
should be defined as 
\[	P= \half (1 - (-1)^F)	\]
for the NS sector, with the ground state having fermion number $F = 0$, 
where 
\[	F = \sum_{n \geq 0} \psi_{-n-1/2} \cdot \psi_{n+1/2}.	\]  
Its most important property is 
\[	\acomm {(-1)^F} {\psi^\mu} = 0.\]
The lowest level that survives this projection consists of 
$8$ massless fields with (spacetime) vector indices.  
It is easy to see that this projection keeps states 
with integral values of $N$ as defined in (\ref {fermionstrength1}).

	In the R sector, we want to project out half of the ground 
states because there are $16$ of them at the start.  This 
can also be accomplished with $(-1)^F$, if we define its 
action on the ground states carefully.  As a representation 
of the $10$-dimensional Clifford algebra, the Ramond ground 
states make a Dirac spinor.  It can be split into 
two irreducible representations of {\it Spin}$(10)$.  They 
are distinguished by their chirality and are mapped into 
each other with any odd power of gamma matrices, i.e.
the zero modes of $\psi^\mu$.  Let us define 
\beq	\label {rf}
	(-1)^F = \pm \gamma^{11} \times (-1)^{F'},
\eeq
where $\gamma^{11}$ is the 10-dimensional chirality operator 
defined as usual in terms of the products of the gamma matrices and 
\[	F' = \sum_{n \geq 1} \psi_{-n} \cdot \psi_{n}	\]
The projection operator $P= \half (1 - (-1)^F)$ will 
project out spinors of either chirality depending on 
the choice of sign in (\ref {rf}).  
Although the overall choice of sign is merely 
a convention, it will become clear in the next section that 
the relative sign between left and right 
movers matters greatly.

	Now let us compute the partition function again.  Inserting 
the projector in the trace, one finds that 
\beqarn	\zNS(P) &=& \Tr_{\mrm{NS}} \, P \, q^{L_0 - 8/24} \\
	&=& \half q^{-1/2} \brak 
		 	{\frac {\prod_{n=1}^{\infty} (1 + q^{n-1/2})^8
		 			- \prod_{n=1}^{\infty} (1 - q^{n-1/2})^8}
		 			{\prod_{n=1}^{\infty} (1-q^n)^8}},
  \\
\zR(P) &=& \Tr_{\mrm{R}} \, P \, q^{L_0 - 8/24} 
	= \half \Tr_{\mrm{R}} q^{L_0 - 8/24}	\\
	&=& \frac {2^4} 2 q^{-1/2} \brak 
		 	{\frac {\prod_{n=1}^{\infty} (1 + q^{n})}
		 			{\prod_{n=1}^{\infty} (1-q^n)}}^8.
\eeqarn
Again, the partition function with completely periodic 
spin structure vanishes.  Amazingly, these two truncated 
partition functions are equal, thanks to Jacobi's \emph 
{aequatio identica satis abstrusa}:
\[	q^{-1/2} \brac {\prod_{n=1}^{\infty} (1 + q^{n-1/2})^8
		 			- \prod_{n=1}^{\infty} (1 - q^{n-1/2})^8}
		 = 2^4 \prod_{n=1}^{\infty} (1 + q^{n})^8		\]

	This result is so remarkable that it is worthwhile to 
understand it in a different light.  In the light-cone gauge, 
we may group the $8$ transverse into $4$ pairs and define
\[
	\Psi^i = (\psi^{2i-1} + \imath \psi^{2i}) / \sqrt 2
\]
and their conjugate $\bar \Psi^i$, for $i = 1, \ldots, 4$.  
Their modes obey the commutation relations 
\[	\acomm {\Psi^i_m} {\Psi^j_n} = 0,~~ \acomm {\Psibar^i_m}
{\Psibar^j_n}=0	\]
\[	\acomm {\Psi^i_m} {\Psibar^{j}_n} = \delta^{i j} \delta_{m+n, 0}.	\]
We may bosonize them: 
\[	\Psi^i = e^{\imath \phi^i_L}.	\]
The (left) momentum for the left moving bosons take values 
in a charge lattice in $\realR^4$, which is different for 
the NS and R sectors.  The $8$ massless states in NS sector 
have charge vectors of the form $(\pm 1,0,0,0),
(0,\pm 1,0,0), ..., (0,0,0,\pm 1)$. 
They are in fact 
the weight vectors\myftnote
	{Weight 
	vectors and conjugacy classes of representation will 
	be defined in the next lecture.}
for the vector representation of $so(8)$.  
The $16$ ground states in the R sector have charges 
$\half (\pm 1, \pm 1, \pm 1, \pm 1)$.  From 
the commutation relations for 
$\Psi$'s and $\Psibar$'s in the R sector, we 
see that their zero modes --- the gamma matrices --- can be expressed 
in terms of fermion creation and annihilation operators.  Using this 
procedure one can explicitly construct the representations of  
Clifford algebra of any dimension.
The $\pm$ sign in each entry of 
the charge vectors for the massless states reflects the occupation 
number of a corresponding fermionic oscillator.
We can define the fermion number 
operator $F$ directly in terms of $\Psi$ as 
\[	F = \Psibar^i_0 \Psi^i_0 +  F'.	\]
Thus the chirality of the massless states is given by the parity 
of the number of minus signs in their charge vectors.  
Not surprisingly, the charge vectors for (anti-)chiral states 
turn out to be the weight vectors for the (conjugate-) spinor 
representation of $so(8)$.  Now there is a \emph {triality} symmetry 
of the Dynkin diagram of $so(8)$ (see fig. 10), which gives isomorphisms 
among the vector, spinor, 
and conjugate spinor representations of $so(8)$.  In fact, these 
isomorphisms extend beyond these three representations and 
hence the massless states. $so(8)$ has 
four conjugacy classes of representations\myftnote
	{See the last footnote.}, 
three of which are represented by 
the vector and two spinor ones respectively.  After GSO 
projection, the charge lattice for the NS sector consists of 
the weight vectors in the vector class, while 
that of the R sector consists of either of the two spinor 
classes.  The equality $\zNS(P) = \zR(P)$ is then 
a consequence of triality.

\myepsf {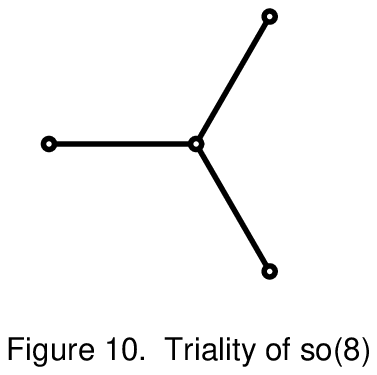}

The triality also allows us to make contact with another 
approach to superstring theory, which is discussed in 
chapter 5 of \cite {GSW}.  
Instead of the spinning particle action 
(\ref {superpt}) which is found to describe a 
spacetime fermion only after quantization, one can define an action with 
spacetime spinor built in and with manifest spacetime 
supersymmetry.  It can be generalized to describe superstrings
(the Green-Schwarz approach).  
To find a relation between the Green-Schwarz approach and
the Neveu-Schwarz-Ramond approach that we are studying
here, let us consider 
the spin field operators which map the NS ground state 
to R ground states:
\beq	\label {gsvar}
	S^\alpha = e^{\frac \imath 2 
			(\pm \phi^1 \pm \phi^2 \pm \phi^3 \pm \phi^4)}, \sepe
		\alpha = (\pm, \pm, \pm, \pm).
\eeq
These spin fields $S^\alpha$ transform as a spinor of $so(8)$ with 
chirality determined by the number of minus signs.  Furthermore 
they all have conformal weight $4 \times \frac 1 8 = \half$ so they 
are also worldsheet spinors.  They are in fact the spinor variables 
used in the Green-Schwarz approach, in the light-cone gauge.  The 
field redefinition (\ref {gsvar}) demonstrates the equivalence 
between Neveu-Schwarz-Ramond and Green-Schwarz superstrings. 

\subsection {Massless Spectrum}

	Now let us examine the massless particles in superstring theory 
for their spacetime meaning.  We will use the language 
of the covariant superconformal gauge, therefore our counting will 
be off-shell.
For NS-NS sector, we clearly get the 
same fields as for bosonic string: the dilaton $\Phi$, the metric 
$G_{\mu\nu}$ and the antisymmetric tensor field $B_{\mu\nu}$.  For the  
NS-R and R-NS sectors, the Ramond parts  
transform as spacetime spinors $\lambda_L$ or $\lambda_R$.  
In fact they are Majorana-Weyl spinors.   
The NS parts are of course vectors, so we have two $10$-dimensional 
Rarita-Schwinger fields.  
The only known way to incorporate such fields consistently
 is to couple them 
to the supergravity current.  They are therefore the 
gravitinos.  So a GSO projected superstring theory contains 
$N=2$ supergravity.  Depending on the choice of the relative sign 
in defining $(-1)^{F_L}$ and $(-1)^{F_R}$, 
we have two inequivalent possibilities, corresponding 
to the relative chirality of the surviving $\lambda_L$ and $\lambda_R$. 
If we choose opposite chiralities, we obtain the type IIA
superstring theory  whose  low energy effective 
theory is the type IIA supergravity.  The type IIA theory is non-chiral 
and can be 
obtained by dimensional reduction from 11-dimensional 
supergravity.  This is the first and simplest evidence for the
relation between type IIA string theory 
and a theory in eleven dimensions, ``\emph{M theory}.''  
M theory is discussed  
by Duff and Schwarz at 
this school.  If we choose the same chirality for both left
and right movers, we obtain the type IIB superstring theory.
The corresponding type IIB supergravity is chiral and 
potentially anomalous.  Cancellation 
of gravitational anomaly in type IIB supergravity 
was shown by Alvarez-Gaum\'e and Witten 
(ref. 20 in \cite{GSW}, Vol 1).

	More novelties come from the R-R sectors.   
Here the massless states 
transform as the products of two spinors.  Contracting them 
with antisymmetrized products of gamma matrices, we see that 
they are related to antisymmetric tensors of rank $0$ to $10$.  
However, because the spinors making the products 
are chiral, not all the possibilities can appear.   
For the type IIA theory, $\lambda_L$ and $\lambda_R$ are
of the opposite chiralities, and we obtain even rank tensors
\[	\Fas 0 \equiv \ \bar \lambda_L \lambda_R, ~~
	{\Fas 2}_{\mu\nu} \equiv \ \bar \lambda_L 
		\gamma_{\mu\nu} \lambda_R, \cdots. \]
On the other hand, the type IIB theory contains odd
rank tensors
\[  {\Fas 1}_\mu \equiv 
		 \bar \lambda_L \gamma_\mu \lambda_R, ~~~
	{\Fas 3}_{\mu\nu\rho} \equiv \ \bar \lambda_L 
		\gamma_{\mu\nu\rho} \lambda_R,
		\cdots. 
\]
Here $\gamma_{\mu_1\ldots\mu_n}$ is the antisymmetrized product of 
$n$ gamma matrices.
Moreover they are not all independent.  There is an important $\gamma$-matrix 
relation: 
\[	{\epsilon_{\mu_1\ldots \mu_n}}^{\rho_{n+1}\ldots\rho_{10}} 
      \gamma_{\rho_{n+1} \ldots \rho_{10}} 
	\sim \gamma^{11} \gamma_{\mu_1\ldots \mu_n}.	\]
Because of the GSO projection, $\Psi_L$ and $\Psi_R$ both 
have definite eigenvalue of $\gamma^{11}$.  Therefore
\beq	\label {duality}
	\Fas n \sim * \Fas {10-n}.
\eeq
In particular, $\Fas 5$ is self-dual.
The students should verify that
the number of independent components of the antisymmetric
tensor fields, taking into account these relations, is 
equal to that of the tensor product of two Majorana-Weyl 
spinors. 
What kind of fields are they? It is not difficult
to show that the massless Dirac equations for 
$\lambda_L$ and $\lambda_R$ are equivalent to   
\[	d^* \Fas n = 0, ~~
	d \Fas n = 0.	\]
They are the equations of motion and Bianchi identities  for 
antisymmetric tensors fields $\Aas {n-1}$ such that 
$\Fas n = d \Aas {n-1}$.  Note that $\Aas {n-1}$ and 
$\Aas {9 - n}$ are related by electric-magnetic 
duality, which exchanges equations of motion and Bianchi 
identities.  The way they arise out of string theory places 
them on equal footing.

	There is also an antisymmetric tensor field $B$ in NS-NS 
sector, but the way it is coupled to the string is very 
different from the R-R fields.  Recall from lecture one 
that the vertex operator for it couples directly to the 
VEV of its potential $B_{\mu\nu}$.  
Its contribution to the string action is just the 
integral of the pullback of $B$ over the worldsheet.
 By analogy with the minimal coupling of the usual 1-form 
potential $A_\mu$ to the worldline of 
a charged point particle, we see that this means 
a string carries unit ``electric'' charge with respect to $B$.  
However, the coupling of R-R fields with string involves 
only the field strength.  This means 
elementary string states cannot carry any charge with respect to the 
R-R fields.  However, it was discovered by Polchinski  
that there are solitonic objects called \emph {D-branes} which do
carry such charges \cite {dbrane}.  These are discussed 
extensively in his lectures at this school.

\subsection {Dilaton and Antisymmetric Tensor Fields}

	The low energy effective action for the NS-NS fields is the same 
as that of the bosonic string:
\[
	S = \frac 1 {2 \kappa ^2} 
		\myI {d^{10}X} 
			\sqrt {-G} e^{-2\Phi} 
			\{ R - \frac 1 {12} H^2 
			+ 4 (\na \Phi)^2 + O({\alpha^\prime}) \},
\]
where $H = dB$.  The variation of $S$ with respect to $B$ gives
\[	 e^{2 \Phi} \na^\mu \left( 
e^{-2\Phi} H_{\mu\nu\rho} \right) = 
	(\na^\mu - 2 \partial^\mu \Phi) H_{\mu\nu\rho} =0.	\]
The origin of the coupling  between $H$ and $\Phi$
can be traced to the way the dilaton couples to 
the string worldsheet, $\sqrt g R \Phi$.  Since 
\(T \sim \frac {1} {\sqrt{g}}
\frac {\delta S} {\delta g^{zz}}\), if the dilaton is
not constant, the energy-momentum tensor $T$ is modified as
\[	T \sim -\half (\pa X)^2 + \pa_\mu \Phi \, \pa_z^2 X^\mu.	\]
The equation of motion for $H$ can then be obtained from the 
Virasoro constraint (\ref {phystatecond}) on physical 
states, which receives the additional contribution from $\Phi$.

	Now let us find out what happens to 
the antisymmetric tensor fields in the 
R-R sector.  The dilaton field also modifies the supercurrent 
as 
\[	G \sim \imath \psi_\mu \pa X^\mu + \psi^\mu \pa_\mu \Phi.\]
As we recall, the zero mode of the super-Virasoro constraint 
yields the massless Dirac equation in the constant dilaton background.  
If the dilaton is not constant, the Dirac operator is modified as 
\[	G_0 \sim \dirac \partial \, - \dirac \pa \Phi = e^\Phi 
\dirac \partial e^{-\Phi}.	\]
Correspondingly, the equations of motion for the R-R fields are 
\[	d^*(e^{-\Phi}\Fas n) = 0, ~~d(e^{-\Phi}\Fas n)=0.	\]
Therefore it is the rescaled fields 
\[	\hFas n \equiv e^{-\Phi}\Fas n	\]
which obey the usual Bianchi identity and equations of motion for 
an antisymmetric tensor.  We can then write $\hFas n = d \hAas {n-1}$ 
and their spacetime action is 
\[	\myI {d^{10} X} \hFas n \wedge * \hFas n	,\]
\emph {without} the usual $e^{-2\Phi}$ factor. 
Thus, we find that the R-R fields do not  couple 
to the dilaton if they are suitably defined.  This is contrary to
the case of the NS-NS $B$ field, for which such rescaling is not
possible. This has far reaching consequences in string dualities, which 
are discussed extensively by other lecturers in this school.

\subsection	{T-Duality}
To end this lecture, let us briefly discuss how the 
T-duality 
$R \to \frac 2 R$ 
 acts on superstring compactified on $M^9 \times S^1$.  Recall 
 from 
the last lecture that this duality involves the isomorphism 
$\pa X_L^9 \bij \pa {X_L^9} '$ and $\pa X_R^9 \bij - \pa {X_R^9} '$.  
This same 
clearly carries over to superstring, but we also have to respect the 
worldsheet supersymmetry.  It is clear that the isomorphism for 
the worldsheet fermions 
should be  
$\psi_L^9 \bij {\psi_L^9} '$ and $\psi_R^9 \bij - {\psi_R^9} '$.  
In particular, 
the zero mode of $\psi^9$ in R sector, which acts as $\gamma^9$ on the
right movers, changes its sign. This means that the relative chirality 
between the left and right movers is flipped.  Therefore 
$R \to \frac 2 R$ maps type II A superstring compactified on a circle of radius $R$ 
to type IIB superstring on a circle of radius $\frac 2 R$.  This is an identification 
of two different types of theories, rather than a gauge symmetry as in the case of 
bosonic string.  What happened is that the operators responsible for 
the enhancement of gauge symmetry, $e^{\pm \imath \sqrt 2 X_L}$, 
are removed by the GSO projection, as are the physical states corresponding to them.

\section {Lecture Four: Heterotic Strings}

	In lecture one we studied the bosonic string which lives 
in $(25+1)$-dimensional spacetime.  It contains only 
spacetime bosons, in particular a tachyon.  In lecture three 
we studied the superstring, which includes spacetime fermions 
in its spectrum, and which, after GSO projection, 
loses the unwanted tachyon and exhibits spacetime 
supersymmetry.  At first sight it seems hardly 
feasible to combine such two drastically 
different theories into 
one without running into disastrous 
inconsistencies.  However, one important property 
of 2d (super)conformal field theories that 
we have used often in the last three lectures is the 
decoupling of left and right movers.  The decoupling 
even extends to 
the zero modes --- momentum and position --- if 
we consider compactification on torus and take 
into account the winding sectors.  In this lecture 
we will exploit this feature 
again and consider a theory with the 
right movers being those of a critical superstring
and the left movers being those of a critical 
bosonic string. This is the
heterotic string of Gross, Harvey, Martinec and Rohm 
(refs. 235, 236 and 237 in \cite{GSW}, Vol 1).  

\subsection	{Marrying Bosonic String and Superstring}

 When we say 
the left movers of the heterotic 
string are those of the bosonic string, 
we mean that they possess the same diffeomorphism 
and Weyl rescaling 
invariance.  The central charge for the ghost 
action is fixed to be $-26$.  Anomaly cancellation 
or equivalently absence of ghosts 
thus requires there to be $26$ left moving bosons in the matter 
sector.  By similar reasoning the right moving sectors 
must consist of 10 matter bosons and fermions.  To have 
genuine target spacetime interpretation as 
a coordinate, a boson must have both left and right movers, 
therefore an ``uncompactified'' heterotic string 
lives in 10 spacetime dimensions.  The additional 16 left 
movers can be thought of as parametrizing an internal 
$16$-dimensional torus. 

	When a theory discriminates between being left and 
right --- when it violates parity invariance --- it is liable 
to incur a gravitational anomaly.  This could be an especially  
acute problem on the $(1+1)$-dimensional worldsheet, where 
the scalars can be chiral and where a chiral fermion and 
its CPT conjugate have the same chirality.  
It would be a disaster for the heterotic string, a 
manifestly left-right asymmetric theory, 
to develop some gravitational anomaly.  
Fortunately this does not happen for the critical heterotic string 
theory we are discussing.  In fact, there is a relation between 
the gravitational anomaly and the Virasoro anomaly.  
Details can be found in \chap 3.2.2--3.2.3 of \cite {GSW}.
Very briefly, from 
(\ref {virasoroope2}) one can deduce that the contributions 
from the left and right movers to the gravitational 
anomaly are proportional to 
their respective central charges.  As shown in the 
reference mentioned above, if and only if they are equal, one 
can introduce local counterterms so that the total gravitational 
anomaly vanishes.  This is certainly true for the critical 
heterotic string theory, where the total central charges 
are 0 for both the left and right movers.

\subsection {Lattice and Gauge Group}

	Let us recall from lecture two that 
an affine Lie algebra $\hat{\Cg}$ 
of level $k$ 
can give rise to spacetime symmetries $\CG$.
When the affine Lie 
currents are present in the physical spectrum for, say, 
the left movers, we can pair it with $\pb X^\mu$ of 
the right movers to make a physical vertex operator.  
Its tree level scattering amplitudes reproduce those of 
a Yang-Mill theory with gauge group $\CG$.  If such vertex 
operators are not in the physical spectrum, say due to GSO 
projection, then $\CG$ cannot be a gauge 
symmetry for the lack of gauge fields.  
However, the worldsheet SCFT still 
possess the symmetry, and the physical states 
and operators fall into representations of 
$\CG$.  So $\CG$ appears 
as a global 
symmetry for the perturbative string theory.  
Now just what kind 
of group $\CG$ can be obtained from 
string theory in this way?

	To answer this question, we need to make a detour to 
the representation theory of Lie groups and algebras.  
We will not focus on the mathematical details but only 
sketch the necessary ideas. 

	Given a finite dimensional 
Lie algebra, we can always find a maximum set of mutually commuting 
generators, the \emph {Cartan subalgebra}.  
We call the commuting generators $H_i$ ($i = 1, \ldots, n$); $n$ is 
the rank of the Lie algebra.  All $H$'s can be simultaneously 
diagonalized in a given representation, and every state 
can therefore be labeled by its eigenvalues for each of the 
$H$'s, which we call charges or quantum numbers.  We may naturally associate
to each set of charges a point in $\realR^n$, a \emph {weight vector}.   
If we plot all of them, they form a lattice in $\realR^n$.  
The reason is that the charges are additive.  
When you multiply two representations, the charge of the product 
of two states is the sum of those of each of them.  As 
every finite dimensional representation can be obtained from  
finite products of a finite set of ``basic'' representations, their 
charge vectors form a lattice, 
the \emph {weight lattice} $\Lambda_W$.  
By the same token, weight vectors of representations 
that can be obtained from products of 
the adjoint representation form 
a sublattice of the weight lattice, 
called the \emph {root lattice} $\Lambda_r$.  
The quotient of the weight lattice by the root lattice 
gives rise to the  
\emph {conjugacy classes} of representations of $\CG$, 
where the conjugation is multiplication with the adjoint 
representation.  
Between the weight and 
root lattices there can be intermediate lattices.  
They and the weight and root lattices, 
are collectively known as 
\emph {Lie algebra lattices}.  
Starting from the Lie algebra $\Cg$, 
one can construct its universal covering Lie group $\CG$.  
The subgroup of $\CG$ whose elements commute with all of $\CG$ 
is known as its \emph {center} $C_\CG$.  
Every element of $C_\CG$ acts nontrivially 
on some representations in the weight lattice, but clearly they 
all act trivially on those in the root lattice.  For representations 
on a Lie algebra lattice, they act as the quotient of $C_\CG$ by some 
subgroup of it.

	Every Lie algebra has an adjoint representation.
Applying the above construction to this particular case, we  
obtain the \emph {Cartan-Weyl} basis: $H_i$ from the Cartan 
subalgebra and the remainder, denoted by $A$, 
that are eigenstates of the $H$, 
\[	\comm {H^i} {A} = a^i A.	\]
Thus each $A$ is associated with a root vector $a^i$ in 
the weight space.  One can show that each root vector is associated 
with only one generator.  

	What kind of construction can realize these  
structures in the context of string theory?  
The additivity of charges 
gives us a hint --- we can represent them as momenta.  
Consider a Lie algebra lattice $\Lambda$ of some 
Lie algebra $\Cg$.  
That the charges take values on the lattice $\Lambda$ 
reminds us of  
compactification over a torus of the same dimension 
as the rank, namely $n$.  Denote the left moving bosons parameterizing 
 the ``torus'' as $\phi^i$.  The Cartan generator $H^i$
is realized by the zero mode of the current $\pa \phi^i_L$,   
as they measure the charges --- momenta.
Therefore this ``torus'' is nothing but the 
maximal Abelian 
subgroup of $\CG$, generated by the $H$'s, known as the 
\emph {maximal torus} of $\CG$.  
Let $\Lambda$ be the charge lattice for the left moving 
bosons.  The momentum  
carried by a state in the lattice is simply 
equal to its weight vector $w$.  It is created 
by the vertex operator $\nod {\exp (\imath w \cdot \phi_L)}$.  
We see now why $\Lambda$ must be a Lie algebra lattice: 
it must contain the adjoint representation so that the 
$A$'s can also be represented as vertex operators.  
Furthermore, those in the 
adjoint should have the same conformal weight 
of $(1, 0)$ as $\pa \phi^i$, 
so they can together form the affine Lie algebra 
(\ref {kac-moody}).  This requires all the generators $A^w$ 
to have $w^2 = 2$.  Lie algebras satisfying this requirement are 
called \emph {simply-laced}.  They are $so(2n)$, $su(n+1)$, 
and $e_n$\myftnote 
	{$e_n$ exits for $n = 6,7, 8$}, 
and the products thereof. The 
$SU(2)$ enhanced symmetry at self-dual radius 
encountered in lecture two is 
their simplest example.  
If $\Lambda$ is the weight lattice, the symmetry group is the 
universal covering group $\CG$.  Otherwise, it is the 
quotient of $\CG$ by some subgroup of $C_\CG$.  To be precise, 
for this construction to 
satisfy the OPE for the affine Lie algebra, we need to introduce 
additional factors known as \emph {cocyles}.
Details can be found 
in \chap 6.4.4--6.4.5 of \cite {GSW}.  Moreover 
we should always remember there is 
a crucial additional requirement from string theory itself --- 
modular invariance.  Therefore the lattice must be 
even and self-dual.

\subsection {$E_8$ Lattice}

	For the heterotic string, the left movers do not 
suffer the GSO projection.  Therefore the vertex operators 
for the non-Abelian generators $A$'s remain in the 
spectrum and we conclude that the theory has 
non-Abelian gauge symmetry with gauge group determined by the 
left components of the lattice.  For the heterotic 
string in $10$ dimensions, the appropriate  
lattice is  $16$-dimensional.  However, it is 
instructive to start with the $8$-dimensional 
even self-dual lattices.

	Let us first state 
some facts about even self-dual lattices $\hat \Lambda$ 
in $(D, D+n)$ spaces.  
It is known mathematically that such objects exist only for 
$n \equiv 0\pmod {8}$.  They are unique up to $O(D, D+n)$ 
isomorphism for $D \neq 0$, and even so for $D=0$ if $n = 8$.  
In $(0, 8)$, the lattice can be chosen to be $\Gamma_{E_8}$, 
generated by 
\[	\bear {ccl}
	e_1 & = & (1, -1, 0, \ldots, 0)	\\
	e_2 & = & (0, 1, -1, 0, \ldots, 0)	\\
	\vdots	& & \\
	e_7 & = & (0, \ldots, 0, 1, -1)	\\
	e_8 & = & ( \half, \half, \ldots, \half ).
	\ear
\]
 The associated \emph {theta function} 
\[	\theta_{\hat \Lambda} (q) \equiv \sum_{p\in \hat \Lambda} q^{p^2}	\]
is invariant under the modular group \slZ. 
The first seven vectors are root vectors of $so(16)$.
The eighth is a weight vector for the chiral spinor
representation of $so(16)$.  Together they generate 
all the weight vectors for the adjoint and  
chiral spinor representations of $so(16)$.  Therefore 
$\Lambda_{E_8}$ is a $so(16)$ Lie algebra lattice.  
Weight vectors for the vector representation take the form 
${} \pm v_i \pm v_j$, where $v_i$ is the 8-vector 
with the $i$-th component 1 and the rest 0.
Those for the chiral spinor representation are 
$({}\pm 1/2, \pm 1/2, \ldots, \pm 1/2)$ with an even number 
of minuses.  
Corresponding to them we have vertex operators 
\beq	\label {ver1}
	e^{{}\pm \imath \phi^a_L \pm \imath \phi^b_L}, 
		\sepe a,b = 1\ldots 8	
\eeq
and 
\beq	\label {ver2}
	e^{\frac \imath 2 ({} \pm \phi^1_L \pm \phi^2_L 
			\cdots \pm \phi^8_L)}. 
\eeq
This suggests us to fermionize these left moving bosons.  
Recall from lecture two that the operators 
\[	\psi^a \equiv e^{\imath \phi^a_L}, \sepe a = 1, \ldots, 8	\]
are 8 complex Weyl (worldsheet) fermions.  We can decompose
them into 16 
Majorana-Weyl fermions:
\[	\Psi^a \equiv \half (\psi^{2a-1} + \imath \psi^{2a}).	\] 
Then the operators in (\ref {ver2}) are just 
the spin fields of $SO(16)$ with a definite chirality.  

	Based on our discussion in lecture two, 
it is easy to write down the partition function 
for these fermions:
\[
	Z_{E_8} = \half q^{-1/3} 
		\brac {\prod_{n=1}^\infty (1+q^{n-1/2})^{16} 
				+ \prod_{n=1}^\infty (1-q^{n-1/2})^{16} 
				+ 2^8 \prod_{n=1}^\infty (1+q^{n})^{16} }.
\]
Here we choose the projection so that the vacuum is \emph 
{not} projected out since the origin is certainly in  
$\Lambda_{E_8}$.  If this were part of a ``compactification'' of 
a bosonic string, its contribution of massless states would be 
from those with weight 1.  From NS sector there are 
$16 \times 15 / 2 = 120$ of them, corresponding to 
the antisymmetrized product of two Majorana-Weyl worldsheet
fermions $\psi^\mu \psi^\nu$.  
 From R sector there are 
$2^8/2 = 128$ of them, corresponding to the R sector vacuum of 
definite $SO(16)$ chirality.

	Which symmetry group would this lattice generate?  
The first thought might be {\it Spin}$(16)$ or 
its quotient by some center.  However, $so(16)$ only has 
120 generators, accounted for the massless states in the 
NS sector.  
The R sector ground states which transform as chiral spinor 
of {\it Spin}$(16)$ also have weight (1,0) and hence correspond to  
affine Lie currents as well.   In fact they enlarge $so(16)$ 
to $E_8$\myftnote 
	{It is customary to denote with $E_8$ both the Lie group 
	and the Lie algebra associated with it.  There is no 
	ambiguity as $E_8$ has only one conjugacy class of 
	representations, which means that there is only one 
	group (i.e. $E_8$) associated with this Lie algebra.}, 
which has $120 + 128 = 248$ generators.  We 
now construct it explicitly.

	Let us start with $so(N)$.  The generators are 
$J^{\mu\nu}$ $=$ $- J^{\nu\mu}$, 
$\mu \neq \nu$ ranging between 1 and $N$.  Their 
commutation relations are well known:
\[	\comm {J^{\mu\nu}} {J^{\rho\sigma}} 
		= \delta^{\mu\sigma} J^{\nu\rho} 
		+ \delta^{\nu\rho} J^{\mu\sigma}
		- \delta^{\mu\rho} J^{\nu\sigma}
		- \delta^{\nu\sigma} J^{\mu\rho}.	\]
To this, let us add a generator $\sigma_\alpha$ with spinor index
$\alpha$.  
Because there exist Majorana-Weyl spinors in (16+0) 
dimension, we may consider Hermitian operators with 
definite chirality.  Their commutation relation 
with the J's, if nonzero, 
must be
\[	\comm {J^{\mu\nu}} {\sigma_\alpha} 
		\sim (\gamma^{\mu\nu})_{\alpha\beta} \sigma_\beta. 
\]
The normalization is fixed by demanding Jacobi identities 
on $[[\sigma, J], J]$.  
The commutators among the $\sigma$'s, after proper 
normalization, must take the form 
\[	\comm {\sigma_\alpha} {\sigma_\beta} 
		= (\gamma^{\mu\nu})_{\alpha\beta} J^{\mu\nu}.	\]
However, one can then check that 
the Jacobi identity for $[[\sigma, \sigma], \sigma]$ 
holds only if  
\[	(\gamma^{\mu\nu})_{\alpha\beta} 
		(\gamma^{\mu\nu})_{\gamma\delta} 
		+ \mbox {cyclic permutation in} (\alpha, \beta, \gamma)
		= 0.	\]
For $so(N)$, this ``Fierz'' type identity holds only for $N$= 8, 9, 16.  \
For $N=8$, it extends $so(8)$ to $so(9)$.  For $N=9$, it extends 
$so(9)$ to $f_4$.  For the relevant case of $N=16$, 
it extends $so(16)$ to $E_8$.  
For more details and 
other interesting facts about $E_8$, the students are referred 
to appendix 6.A of \cite {GSW}.

\subsection {$E_8 \times E_8$ and {\it Spin}$(32)/\intZ_2$}

	Now let us consider $16$-dimensional self-dual even 
lattices.  Mathematically, it is known that 
there are two of them 
up to $SO(16)$ rotations.  One of them is simply the 
direct product of 2 copies of $\Lambda_{E_8}$.  Its 
generators, in one-to-one correspondence with weight 
one vertex operators, are simply generators of either 
of the two $E_8$'s.  The associated partition function 
\[	Z_{E_8 \times E_8} = Z_{E_8}^2. \]

	But there is another lattice, unrelated to the 
one above by any $SO(16)$ rotation yet equally simple to 
describe.  It is generated by 
\[	{}\pm w_i \pm w_j,~~ i \neq j,\]
where $w_i$ is now a $\realR^{16}$ vector with the $i$-th components 
1 and the rest 0, and 
\[	({}\pm 1/2, \pm 1/2, \ldots, \pm 1/2),	\]
with an even number of minuses.  By analogy with 
$\Lambda_{E_8}$, it contains the root vectors 
of $so(32)$ and the weight vectors of its chiral spinor 
representation.  It is the $so(32)$ Lie algebra lattice.  
The difference between this and the last case is that the 
chiral spin fields now have weight $(2, 0)$ so 
do not form the currents.  The weight $(1, 0)$ 
operators all correspond to the roots of $so(32)$.  
The lattice does not include the vector and anti-chiral 
spinor representations.  So the gauge group is not 
quite {\it Spin}$(32)$, but rather its quotient by a 
$\intZ_2$ subgroup of its $\intZ_2 \times \intZ_2$ center.  
It is usually written as {\it Spin}$(32)/\intZ_2$ to 
distinguish it from $SO(32)$\myftnote 
	{$SO(32)$ is the quotient of 
	{\it Spin}$(32)$ by the other $\intZ_2$ in its 
	$\intZ_2 \times \intZ_2$ center.  It would have been the 
	gauge group if the Lie algebra lattice had included 
	both the adjoint and the vector representations 
	but neither of the two spinor 
	representations.}.
It is simple to check that $so(32)$ has the same number 
of generators as $E_8 \times E_8$, namely 496.
We can also calculate the partition function
\beqarn
 Z_{SO(32)/\intZ_2} = &\half q^{-2/3} &
 \left\{ \prod_{n=1}^\infty (1+q^{n-1/2})^{32}	\right. \nono
 && \left. + \prod_{n=1}^\infty (1-q^{n-1/2})^{32}
 + 2^{16} q^2 \prod_{n=1}^\infty (1+q^n)^{32} \right\}.
\eeqarn
By using Jacobi's abstruse identity, introduced in lecture three, 
it is easy to show that\myftnote
	{This is not a coincidence.  Mathematically it is known 
	that there is a unique modular form of modular weight 8.} 
\[	Z_{SO(32)/\intZ_2} = Z_{E_8 \times E_8} = Z_{E_8}^2.	\]

\subsection {Particle Spectrum}

	We now study the low lying particle spectrum for the 
two heterotic string theories in 10 dimensions.  The procedure 
for the left movers is identical to that of bosonic string;  
for the right movers it is identical to that of the type II string.  
Therefore we will be very brief.

	For the right movers, the mass shell condition is 
$L_0 = 1/2$ or 
\[	m^2 = -p^2 = 2 \tilde N \]
where $p$ is the $10$-dimensional spacetime momentum, and 
$\tilde N$ is the measure of oscillator excitation defined as 
in (\ref {fermionstrength1}) and (\ref {fermionstrength2}).
The ground state is projected out by GSO projection, 
so the lowest lying physical states are either 
\[	e_\mu \tilde {\psi}_{-1/2}^\mu \ket {k}, \sepe 
		\mu = 0, \ldots, 9\]
in the NS sector, satisfying the massless Klein-Gordon equation
\[	k^2 = 0,~~ k \cdot e =0,\]
or the ground states in the R sector with 
definite chirality:	
\[	\xi^\alpha \ket {k}_\alpha,	\]
satisfying the massless Dirac equation
\[	\dirac k \xi = 0.\]

	For the left movers, the mass-shell condition 
is $L_0 = 1$ or 
\[	m^2 = -p^2 = 2 (N - 1) + p_L^2,\]
where $p_L$ is 
the internal momentum living on the $16$-dimensional 
even self-dual lattice, and $N$ the measure 
of left moving oscillator excitation as defined in 
(\ref {bosonstrength}).  The ground 
state is 
\[	\ket {k}, \come k^2 = 2.\]
Because of the left and right asymmetry, the level matching 
condition for the heterotic string is modified\myftnote
	{One can again understand this by looking at the 
	partition function.  The integration over the twist moduli 
	$\im \tau$ enforces the level matching condition 
	(\ref {hetlevelmatch}).  The constant $1$ and 
	$1/2$ originate 
	from the different central charges of the 
	left and right movers in the light-cone gauge.}:
\beq	\label {hetlevelmatch}
	N + p_L^2/2 = \tilde N + 1.
\eeq
Note that this means $p_L^2$ and hence the internal 
lattice must be even.  Note also that for $N=0$, 
$p_L^2$ must be at least 1.  
This means that although the left movers have no GSO projection, 
the tachyon is still projected out.  
The first excited states are massless.  
They include the usual
\[	e^\mu \alpha_{-1}^\mu \ket {k} \sepe \mu = 0, \ldots 9	
	k \cdot e = 0\]
and contribution from the internal bosons:
\[	J_{-1}^a \ket {k},	\]
$J_n^a$ being the Fourier modes of the current $J^a$.

	Putting the left and right movers together, 
the massless spectra of the heterotic strings 
include the usual spacetime bosons 
$G_{\mu\nu}$, $B_{\mu\nu}$, and $\Phi$ coming 
from
\[	\alpha^\mu_{-1} 
	\tilde {\psi}^\nu_{-1/2} \ket {k},	\]
and spacetime fermions --- gravitinos --- coming from 
\[	\alpha^\mu_{-1} \ket {k}_\alpha. \]
These are similar to what one would get from NS-NS and NS-R 
sectors of superstring, but the additional 
16 left moving bosons or, 
equivalently, 32 left handed fermions give rise to something quite 
new.  We have now gauge fields from 
\[	J^a_{-1} \tilde {\psi}^\mu_{-1/2}\ket {k},	\]
and gauginos from 
\[	J^a_{-1} \ket {k}_\alpha	.\]
Therefore the low energy approximation to a heterotic 
string theory would be a theory with $N=1$ supergravity 
and $N=1$ super-Yang-Mills.  
It is anomaly free only when the gauge 
symmetry algebra is $E_8\times E_8$, $so(32)$, 
$u(1)^{248} \times E_8$ or $u(1)^{496}$.  We have 
thus explained the ``existence'' of  
the first two as being low energy approximation 
to the two heterotic string theories.

\subsection {Narain Compactification}

	Recall that in lecture two, we considered
generalized compactification over $T^D$  
by letting the internal left and right momenta to live on a 
$(D+D)$-dimensional lattice $\hat \Lambda$.  The requirement of 
modular invariance then places stringent restrictions 
on $\hat \Lambda$.  This construction can be carried over 
for the heterotic string, in which case the 
left moving bosons have $16$ more ``dimensions'' than the 
right moving ones.  Thus in 
toroidal compactification down to $10-D$ dimensions, 
the left and right momenta lives on a $(16+2D)$-dimensional 
lattice $\hat \Lambda_H$.  Modular invariance again 
requires $\hat \Lambda_H$ to be even and self-dual with 
respect to a metric of signature $(16+D, D)$.  Such a 
$\hat \Lambda_H$ 
is known as \emph {Narain lattice} (ref. 340 in \cite{GSW}, Vol 1).

	Following the discussion earlier, the non-Abelian 
gauge symmetry of the 
compactified heterotic theory is determined 
by the special points  
in the lattice of the form 
$(p_L, 0)$ with $p_L^2 = 2$, and the global 
symmetry determined by those with charge vector 
$(0, p_R)$ with $p_R^2 = 2$.  
Generically, there will be no points like those, 
and the gauge symmetry of the theory 
is Abelian $U(1)^{16+D} \times U(1)^D$.  
The $U(1)^{D} \times U(1)^D$ are just the Kaluza-Klein 
gauge fields.  The $U(1)^{16}$ is what remains of 
the original gauge symmetry of the heterotic string.  The 
breaking of the gauge symmetry
$E_8 \times E_8$ or {\it Spin}$(32)/{\Bbb Z}_2$ down to products 
of $U(1)$ is achieved by turning on \emph {Wilson lines}, which 
we will discuss presently.  Let us note, however,  that there 
are also nongeneric lattices where such special points do exist.  
The self-dual radius is again the simplest example.  
As in that case, we would have an enhancement of 
gauge and/or global symmetries.  
The existence of such points plays an important
role in understanding the string-string duality
between the
heterotic string on $T^4$ and the type II A string
on $K_3$.  It is discussed in 
Aspinwall's lectures in this school.

	The discussion in lecture two on the moduli spaces for toroidal 
compactification 
can be carried over to the present case.  As expected, they are 
$$
 O(D+16, D; \intZ)\backslash 
 O(D+16, D) /
 O(D+16) \times O(D)	
$$
for $D > 0$.
By arguments similar to those given in lecture two, these 
\emph {Narain moduli} are VEV's for the massless fields 
$\pa X^M \pb X^N$ and $\pa \phi^i \pb X^N$, where $M$ are 
indices tangent to the compactification torus $T^D$ and 
$i$ are labels in the Cartan subalgebra of 
either {\it Spin}$(32)/{\Bbb Z}_2$ or $E_8 \times E_8$.
The first type are just the familiar Kaluza-Klein scalars 
$G_{MN}$, $B_{MN}$.  The latter 
are components $A^i_M$ 
of the gauge fields in the Cartan.  
For $D = 0$, the moduli space consists of two discrete points, 
corresponding 
to $E_8 \times E_8$ and {\it Spin}$(32)/{\Bbb Z}_2$. 
However, as mentioned earlier, 
for $D>0$ the moduli space is connected.  This has the interesting 
implication that one can continuously interpolate between the 
two heterotic string theories compactified over $T^D$.  
We now sketch one such interpolation for compactification 
on $S^1$.  Starting 
with $SO(32)$, we give some constant VEV's to $A_9^i$ in 
the Cartan.  This 
is known as ``turning on the Wilson line'' around $S^1$.  
It is so called because it 
lets the Wilson loop around $S^1$, i.e. 
the path ordered exponential 
\[	P \exp^{\paren {i \int_{S^1} A_9  dx^9}},	\]
develop a nontrivial VEV, which can be chosen to
break $SO(32)$ down to 
$SO(16) \times SO(16)$.  After an appropriate 
$O(17, 1;\intZ)$ T-duality transformation, 
it becomes a Wilson line 
configuration for the $E_8 \times E_8$ heterotic 
string compactified on $S^1$.

\begin{exercise}{}
	Another way to obtain gauge symmetry in string 
theory is to consider open strings.  This subject 
is discussed extensively in Polchinski's 
lectures.  For this exercise, reconsider bosonic 
string on a worldsheet $\Sigma$ 
with a \emph {boundary} $\pa\Sigma$.  To solve the Cauchy 
problem, one must impose boundary conditions along 
$\pa\Sigma$.  This leads to new constraints on the phase 
space.  Repeat the classical and quantum analysis 
of lecture one for this case, assuming the Neumann 
boundary condition
\[
	\left. \pa_{normal} X = 0 \right|_{\pa\Sigma}
\] 
for all $X$'s.  Find the 
Virasoro constraints and 
determine the massless spectrum.  
What happens if instead we use Dirichlet boundary 
condition 
\[
	\left. \pa_{tangential} X = 0 \right|_{\pa\Sigma}
\] 
for some $X$'s?
\end{exercise}

\section {Lecture Five: Orbifold Compactifications}

	Although simple and interesting, toroidal 
compactifications cannot give rise to realistic 
theories because 
they have a rather large number of 
unbroken spacetime supersymmetries
for the uncompactified spacetime.
To see this, consider the compactification 
over $T^6$.  Both heterotic string theories have 
$N=1$
spacetime supersymmetry in 10 dimensions, corresponding to  
$2^4 = 16$ real components of 
supercharges forming a constant Majorana-Weyl spinor in 
$(9+1)$-dimensions.  
Because $T^6$ is flat, all of them survive 
as unbroken supersymmetry for $M^4$.  $N=1$ 
supersymmetry in $M^4$
has $4$ real components of supercharge.  
Thus the heterotic string compactified on $T^6$ 
gives rise to an $N=4$ theory in 4 dimensions.  The number 
of supersymmetries is doubled for type II theories, 
because they start with $N=2$ in 10 dimensions.

	To obtain realistic models 
one has to consider compactifications on more 
complicated manifolds known as 
Calabi-Yau spaces or more general 
superconformal field theories as the internal 
part.  These are discussed extensively 
in Greene's lectures at this school.  
Here we will discuss the simplest type 
of Calabi-Yau spaces, known 
as \emph {orbifolds} \cite{orbifold}. 

\subsection {$S^1 / {\Bbb Z}_2$}

	This is the simplest illustration of the idea of 
orbifold compactification.  
As you recall from lecture two, $T^1 \sim S^1$ 
 can be defined as the quotient of $\realR^1$ by $2\pi R \intZ$.  
Now let us consider a further $\intZ_2$ equivalence relation:
\[	 X \sim -X.	\]
This defines the quotient $S^1 / \intZ_2$.  What does the 
resulting space look like?  To find out, note that it has two fixed points: 
0 and $\pi R$.  The latter is a fixed point because 
$-\pi R \sim \pi R$ on the $S^1$.  
$S^1 / \intZ_2$ therefore looks like a line segment 
(fig. 11).

\myepsf {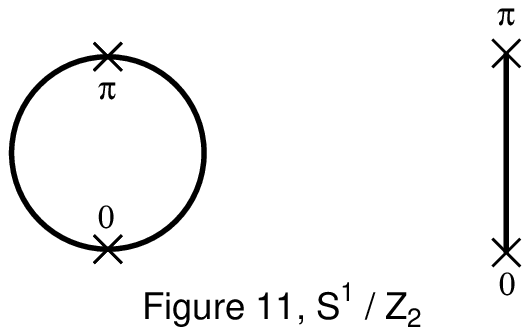}

	Recall that in toroidal 
compactifications, requiring the 
spacetime wavefunction to be 
single valued results in the quantization of 
center of mass momentum.  We could alternatively 
say that we project out all the states which 
are not invariant under the equivalence 
relation defining the torus (\ref {torusdef1}) 
with the operator
\[	\sum_{\Delta X \in \Lambda} e^{ip \cdot \Delta X}\]
where $e^{ip \cdot \Delta X}$ is the operator 
that performs a translation by the lattice vector 
$\Delta X$.  It is clear that  
this operator is simply a periodic delta function in momentum
space singling out the correctly quantized momenta.
Similarly, for orbifold compactification we should project 
out states which are not invariant under the $\intZ_2$ 
operation with the projection operator 
\[	P =  (1 + \Omega)/2 \]
where $\Omega$ is the operator that perform 
the appropriate $\intZ_2$ on $X$:
\[	\Omega^{-1} X(z, \zb) \Omega = -X(z, \zb).\]  
This is very similar to the action of $(-1)^F$ 
introduced in lecture two, so it
is easy to see that the partition function is 
\beqar	\label {untwistedparti}
	\lefteqn {Z_u \equiv \Tr \, \paren {P \, 
	q^{L_0 - 1/24} \qbar^{\tilde {L}_0 - 1/24} } }\nono
	&=& \half \brac {\frac 1 {\abs{\eta}^2} 
			\sum_{p, \tilde p} q^{p^2/2} \qbar^{\tilde {p}^2/2}
			+ \frac 1 {\abs{q^{1/24} \prod_n (1+q^n)}^2}}. 
\eeqar 

	There is an immediate problem with this partition function.  
We know the first term in (\ref {untwistedparti}) is 
modular invariant, because it is simply the internal part of 
the partition 
function for the string compactified on $S^1$, derived in lecture two.  
However, it can be checked that the second 
part is not modular invariant.  In fact it is easy to figure
out the modular transformation property of the second term
since $q^{1/24} \prod_n (1+q^n)$ is exactly the partition function 
of the free fermion studied in lecture two. 
Under the modular transformation $S$ 
\[	\abs{q^{1/24} \prod_n (1+q^n)}^2	\]
becomes 
\[	\half \abs{q^{-1/48} \prod_n (1-q^{n-1/2})}^2, \]
which then becomes 
\[	\half \abs{q^{-1/48} \prod_n (1+q^{n-1/2})}^2	\]
after the $T$ transformation.  Therefore we must include all of them 
in the modular invariant partition function
\beqar	\label {twistedparti}
	Z &= \half q^{- 1/24} \qbar^{ - 1/24} 	
	& \left\{  \frac {\sum_{p, \tilde p} q^{p^2/2} \qbar^{\tilde {p}^2/2}}
			{\abs{q^{1/24} \prod_n (1-q^n)}^2}
			+ \frac 1 {\abs{q^{1/24} \prod_n (1+q^n)}^2}	\right. \nono
		& &	\left. + \frac {2 q^{ 1/16} \qbar^{ 1/16}}
					{\abs {\prod_n (1-q^{n-1/2})}^2}
			+ \frac {2 q^{ 1/16} \qbar^{ 1/16}}
					{\abs {\prod_n (1+q^{n-1/2})}^2}
					\right\} .
\eeqar

	What is the meaning of the last two terms?  Recall again 
the case of toroidal compactification.  There  not only do we 
quantize the center-of-mass momentum to ensure the 
single-valuedness of the wavefunction, but we 
also have to take into account the winding sectors, which represent 
strings wrapping around nontrivial loops on the torus: 
\[	X(\sigma^1+2\pi, \sigma^2) = 
		X(\sigma^1, \sigma^2) + 2\pi m R.\]
On $S^1/\intZ_2$ there are more sectors due to 
the identification $X \sim -X$.  
We should consider \emph {twisted sectors}, which 
correspond to 
\[	X(\sigma^1+2\pi, \sigma^2) = 
		- X(\sigma^1, \sigma^2) + 2\pi m R.\]
The minus sign in this boundary condition 
requires that the modding of $X$  be half integral.
\[ X = x + 
		\sum_{n \in \intZ + 1/2} \frac \imath n 
		(\alpha_n e^{-nz} + \tilde \alpha_n e^{-n\zb}).	\]
We cannot have nonzero momentum or winding number here since 
they are not consistent with the anti-periodic boundary condition.
The boundary condition also restricts $x$ to be 
$0$ or $\pi R$.  
Therefore there are two twisted sectors, each centering 
on a fixed point of the 
$\intZ_2$ action on $S^1$.  This is a general 
feature of orbifold compactification.

	The additional terms in the partition function 
can now be understood as contribution from the two 
twisted sectors.  They both give the same contribution 
\beqarn
	\lefteqn {\Tr_{twisted} \: \{ P \, 
	q^{L_0 - 1/24} \qbar^{\tilde {L}_0 - 1/24} \} } 
			\nono
	&&= \half q^{- 1/24} \qbar^{ - 1/24} 
	\brac { \frac { q^{ 1/16} \qbar^{ 1/16}}
					{\abs {\prod_n (1-q^{n-1/2})}^2}
			+ \frac { q^{ 1/16} \qbar^{ 1/16}}
					{\abs {\prod_n (1+q^{n-1/2})}^2}}.
\eeqarn
Note that the formula (\ref{twistedparti}) contains
the factor $2$, reflecting the fact that there are
two fixed points of $\intZ_2$. 
Modular transformation mixes the partition 
function for twisted 
and untwisted sectors, with or without the insertion 
of the operator $\Omega$, in exactly the same fashion 
it mixes different spin structures as discussed in 
lecture two.

	Recall that in the free fermion theory, 
the ground state of the periodic, Ramond, sector 
has a higher energy relative to the anti-periodic, 
Neveu-Schwarz sector. 
For the bosonic orbifold theory, however, 
the ground state of the anti-periodic, i.e. twisted sector,
has a higher eigenvalue of $L_0$ and $\tilde {L}_0$.  
Its weight is (1/16, 1/16)\myftnote
	{This can be obtained by computing the OPE 
        of the energy-momentum tensor 
	with a twist field, which generates the twisted
        boundary condition of $X$, or by the
	$\zeta$-function regularization. Here
	we derive it by requiring modular invariance.}
per twisted coordinate.
This is the same as that of the R-R ground states 
per real fermion.

\subsection {$T^4/\intZ_2$}

Now let us consider an only slightly more involved 
example which is nonetheless already 
a limiting case of Calabi-Yau 
compactification.  
The compact manifold is now $T^4/\intZ_2 \sim (S^1)^4/\intZ_2$, 
where the 
$\intZ_2$ acts on each of the four $S^1$ as in the last 
example:
\[	X^i \to - X^i, \come i = 1, \ldots, 4.\]
As each $S^1$ has 2 fixed points, on this 
orbifold there are $2^4 = 16$ fixed points.  
An analysis similar to the one given above shows 
there is a twisted sector associated with 
each of them.  The weight of their ground state 
is 
\[	\left(\quat, \quat
\right) = 4 \times \left(\frac 1 {16}, \frac 1 {16}
\right).	\]

Since we want to discuss superstring compactified on 
this orbifold, we should include the worldsheet 
fermion $\psi$'s as well.  They transform as 
tangent vectors in spacetime.  Now the $\intZ_2$ map 
clearly acts on the tangent space as well, as it 
reverses spacetime direction:
\[	\psi^i \to - \psi^i.	\]
In fact this is also required by the  
superconformal invariance, which mixes between 
$X^i$ and $\psi^i$.  As the $\psi$'s already 
have periodic and anti-periodic boundary 
condition, 
the $\intZ_2$ action merely exchanges their 
assignment to R and NS sectors respectively.  
Previously we saw that 
each $\psi^i$ increases the conformal
weight of the ground state by 
${1 \over 16}$ when going from the NS to
the R sector. Thus, in the twisted
sector, the fermions should contribute
$4 \times {1 \over 16} = {1 \over 4}$ 
to the conformal weight. The total conformal weight of
the twisted sector is then  (1/2, 1/2).  
In particular, they correspond 
to massless states in the physical spectrum of 
type II superstring. 

In fact each fixed point gives rise to 4 massless
scalar fields in the uncompactified $(5+1)$ dimensions. 
In order to change the boundary condition of the
fermions, we may bosonize the 4 fermions into 2 bosons
$\phi^1$ and $\phi^2$, and consider the spin operators 
\[ \sigma_{\pm\pm} = e^{\pm {\imath \over 2} \phi^1
\pm {\imath \over 2} \phi^2} . \]
The GSO projection forces the number of minuses to
be even, so there are 2 choices. Since the left and
the right movers can have different choices, there are
$2 \times 2=4$ ways to change the boundary condition 
of the fermions. Since there are $16$ fixed points, 
the type II superstring on $T^4/\intZ_2$ gives 
$4 \times 16 = 64$ massless scalar fields from
the twisted sector. 

In addition, there are $4 \times 4=16$ massless
scalars coming from the untwisted sector. 
They are constant modes of the metric $G_{ij}$
and the NS-NS $B_{ij}$ ($i,j=1,...,4$) and correspond
to the Narain moduli of $T^4$.  
In fact, the $64$ scalars from the twisted sector 
share a similar geometric interpretation.  
They are so-called \emph {blow-up} modes, and 
their VEV's deform and resolve the orbifold 
singularity at the fixed points.  
When these singularities are fully resolved, 
one recovers a smooth Calabi-Yau manifold known 
as $K_3$.  Combining the twisted and the untwisted
sectors together,  
the moduli space of type II string 
compactification over $K_3$ is  
$16+64 = 80$.  This is the same as that 
of the heterotic string compactified over 
$T^4$ since $4 \times (16 + 4) = 80$.  
This is not a mere coincidence, and 
its deeper reason will be uncovered during the
school.

We hope you have acquired the necessary knowledge to cope with
the more advanced lectures in this school. {\it Bon Voyage}!

\section*{Acknowledgments}

We would like to thank the school organizer K.T. Mahanthappa
and the program director B. Greene for their beautiful planning and
organization of the school. We thank J.~Harvey and 
J.~Polchinski for their useful suggestions on the outline 
of the course and for J.~Feng and M.~Peskin for having
a preliminary draft of their forthcoming book available to us. 
We are also grateful to C-S.~Chu, Y.~Oz  and H.~Steinacker for reading 
the draft. This work was supported in part by the National Science
Foundation under grants PHY-951497 and in part
by the Director, Office of Energy Research, Office of High Energy
and Nuclear Physics of the U.S. Department of Energy under Contract
DE-AC03-76SF00098.  Z.Y. is supported in part by
a Graduate Research Fellowship
of the U.S. Department of Education.

\end{document}